\documentclass[]{myaa}
\usepackage{natbib}
\usepackage{hyperref}
\usepackage{txfonts}
\usepackage{graphicx}
\usepackage{psfig}
\usepackage{aalongtable}
\usepackage{balance}
\begin{document}
\idline{0}{00}
\let\year=2
\def\teff{$T\rm_{eff}$ }
\def\kms {$\mathrm{km\, s^{-1}}$ }
\def\vsini {$\mathrm{v\,sin\,i$}}

\title{{Sulphur abundance in Galactic stars}
\thanks {based on observations collected at ESO in
programmes: 056.E-0665, 59.E-0350,  62.L-0654 and  165.L-0263
}}

\author{
E. Caffau     \inst{1} \and
P. Bonifacio  \inst{2} \and
R. Faraggiana \inst{3} \and
P. Fran\c cois \inst{4} \and
R.G. Gratton \inst{5}
\and
M.~Barbieri \inst{5}
}

\institute{
Liceo L. e S.P.P. S. Pietro al Natisone, annesso al
Convitto Nazionale ``Paolo Diacono'',
Piazzale Chiarottini 8, Cividale del Friuli (Udine), Italy\\
   \email{elcaffau@yahoo.it}
\and
Istituto Nazionale di Astrofisica - Osservatorio Astronomico di
Trieste,
    Via Tiepolo 11, I-34131
             Trieste, Italy\\
   \email {bonifaci@ts.astro.it}
\and
Dipartimento di Astronomia, Universit\`a degli Studi di Trieste
Trieste,
    Via Tiepolo 11, I-34131
             Trieste, Italy\\
   \email{faraggiana@ts.astro.it}
\and
Observatoire de Paris, 64 Avenue de l'Observatoire, F-75014 Paris, France
\and
Istituto Nazionale di Astrofisica - Osservatorio Astronomico di
Padova, Vicolo dell'Osservatorio 5, I-35122 Padova, Italy
}

\authorrunning{Caffau et al.}
\titlerunning{Sulphur abundance in Galactic stars}
\offprints{E. Caffau}
\date{Received ...; Accepted ...}

\abstract{We investigate sulphur abundance in 74 Galactic stars
by using high resolution spectra obtained at ESO VLT and NTT
telescopes. For the first time the abundances are derived, 
where possible,
from three optical multiplets: Mult. 1, 6, and 8.
By combining our own measurements with data in the 
literature we assemble a 
sample of 253  stars in the metallicity range
$\rm -3.2 \la [Fe/H] \la +0.5$.
Two important features, which could hardly be
detected in smaller samples, are obvious
from this large sample: 1) a sizeable 
scatter in [S/Fe] ratios around [Fe/H]$\sim -1$ ; 2) at low metallicities
we observe stars with [S/Fe]$\sim 0.4$,
as well as stars with higher [S/Fe] ratios. The latter 
do not seem to be kinematically
different from the former ones.
Whether the latter finding stems from a
distinct population of metal-poor stars or simply from
an increased scatter in sulphur abundances remains an open
question.

}

\maketitle

\section{Introduction}

Sulphur has been a long-neglected element in the 
study of Galactic chemical evolution;
after
the pioneering works of \citet{clegg} 
for stars with 
[Fe/H] $\geq -1$  and
\citet{fran87,fran88}
for the most metal-poor stars, nothing 
was published until recent years.
This light interest was largely due
to the difficulty of measuring S abundances in
stars, as detailed in the next sections,
but also to the fact that since abundances
of nearby $\alpha-$elements Si and Ca 
were readily available from the analysis of
stellar spectra, it was felt
that the additional insight into
nucleosynthesis and chemical evolution
which could be derived from 
sulphur abundances was not worth the great effort
necessary to measure them.
From the nucleosynthetic 
point of view, sulphur is made by 
oxygen burning, like Si and Ca, either in a central
burning
phase, convective shell, or explosive phase according to
\citet{limongi}.
This is a strong reason why Si, S, and Ca
are expected to 
vary in lockstep with chemical evolution.

However, in recent years the study of chemical
evolution in external galaxies has gained impetus.
For this purpose, the more readily available objects 
are Blue Compact galaxies (BCGs, \citealt{garnett,tp89}) through analysis
of the emission line spectra, and Damped Ly$-\alpha$
systems (DLAs, \citealt{centurion})
through the analysis of resonance absorption
lines. In both groups of objects sulphur, is relatively 
easy to measure. And
for both groups of objects  a gaseous 
component of the galaxy is measured,  being
always aware of possible corrections
to the measured abundances for the fraction
of elements which are locked in dust grains
(depletions). From this point of view sulphur
is  a very convenient element to use
because
it is known from the study of the Galactic
interstellar medium that sulphur is a volatile
element, i.e. it forms no dust.
It has thus become very interesting
to provide a solid Galactic reference
for sulphur abundances, which may be directly
compared to measures in external galaxies.

Thus in  recent years there have been 
a number of studies with this
goal \citep{israelian,chen02,takeda,chen03,nissen,ryde,ecuvillon},
leading to a somewhat controversial picture.
Some studies claim that sulphur behaves
exactly like  silicon and other
$\alpha$ elements \citep{chen02,chen03,nissen,ryde},
which display  a ``plateau'' in their ratios to iron, while
other studies seem to favour a linear
increase of [S/Fe] ratios with decreasing
metallicities \citep{israelian,takeda}.

To shed new light on the problem of the evolution
of sulphur in the Galaxy,  
we analyse high resolution
spectra of Galactic stars that were collected
in the course of several of our
observational programmes, as well as spectra
retrieved from the ESO archive.

\section{Observational Data}

The sample of stars was obtained by combining observations made
at ESO with the NTT and VLT telescopes (see Table \ref{data}). 
In this table we report
the S/N ratio at  670 nm,
when available, otherwise  we report the
S/N ratio at 
870 nm.
The spectra  of most stars  were already used
for different investigations and the observational
details published in other papers.

\subsection {NTT spectra}

NTT data were obtained in three different runs,
programmes  56.E-0665, 59.E-0350, and  62.L-0654. 
The data of programme 56.E-0665 were retrieved
from the ESO archive. The observations were obtained  in January 1996, 
and abundances from several elements derived from these
spectra were presented by
\citet{NS}.
The   EMMI instrument  configuration
used was  echelle \# 14  and grism \# 3 as a cross-disperser,
and the resulting resolution was R$\sim 60000$.  
These spectra are marked as ESONTTB in the last column of  Table \ref{data}.
We performed the observations of programmes 59.E-0350 and 62.L-0654
in August 1997 and March 1999,
respectively. 
In both cases the EMMI instrument was 
used with echelle \# 14 and grism \# 6 as cross-disperser;
also in this case the  resolution was  R$\sim $ 60000.
These spectra are marked as ESO-NTT in the last column of  Table \ref{data}.
The different cross-dispersers and central wavelengths
result in different spectral coverage, nevertheless 
for all NTT data the 
only  \ion{S}{i} lines available were those of Mult. 8.

\subsection {VLT spectra}

All  VLT data were obtained in the course
of the Large Programme 165.L-0263 (P.I. R. G. Gratton),
which used 
the UVES spectrograph mounted on the Kueyen-VLT 
8.2 m telescope at Paranal. 
For most observations the slit width was $1''$
yielding a
resolution  R$\sim$ 43000, and occasionally, due to variable seeing 
conditions, slightly smaller or larger slit widths were used.
The data for all the stars in the present paper were also
used by \citet{gratton}.
All  observations were taken with dichroic \# 2 
and grating \# 4 as cross disperser.
Only the red arm spectra are discussed here;
in the first run of June 2000, the central 
wavelength of the red arm was set to 700 nm, which allows
one to cover the range 520--890 nm with a small
gap between 700 nm and 705 nm, corresponding
to the gap in the red arm CCD mosaic. Therefore
for this run, only the \ion{S}{i} lines
of Mult. 6 and 8 are  available.
In all subsequent runs the central wavelength
was set at 750 nm, thus allowing coverage of the range
575--931 nm, with a gap of about 5 nm around 750 nm.
Therefore for this set of data all three \ion{S}{i} multiplets
discussed in this paper, i.e. Mult. 1, Mult. 6 and 
Mult. 8, are available.

\section {Atmospheric parameters}

Most stars studied here are in common with 
\citet{gratton};
we adopted their values of the atmospheric parameters \teff, log g, 
and [Fe/H],
in order to compare [S/Fe] to their value of [Mg/Fe].
For the other stars, we used IRFM temperatures from Alonso
(private communication)
and surface gravities derived from the Hipparcos parallax
(Eq. 5 of \citealt{gratton}),
whenever available, or  from the
\ion{Fe}{i} /\ion{Fe}{ii} 
ionization balance, otherwise.
For the remaining stars 
we used the atmospheric parameters of \citet{NS}.
The errors on \teff are about 50 K according to \citet{gratton}
and 
about 40 K  according to \citet{NS}.
The error on IRFM temperatures is larger,  on the order of 100 K.
The internal error  on log g 
from Hipparcos-based parallaxes is on the order of 
0.1 dex and is dominated by the error on the parallax.
The error on surface gravities based on 
\ion{Fe}{i}/\ion{Fe}{ii} 
ionization balance 
is again on the order of 0.1 dex \citep{NS}.

The choice of the microturbulence  ($\xi$)  does not affect  the S-abundance
determination from the weak lines of Mult. 6 and 8, but may be important
for the stronger lines of Mult. 1 for the less metal poor stars (see Sect. 6).
To analyse the lines
of all multiplets, we adopted the ${\xi}$  value in \citet{gratton}.
For multiplets 6 and 8 we adopted the ${\xi}= 1$ \kms when 
\citet{gratton} was not available.

\section{ The \ion{S}{i} spectrum}

Preliminary work consisted in  selecting lines which can be 
measured on spectra of metal-poor  stars and retrieving
the most reliable log gf values.

\begin{figure*}
\psfig{figure=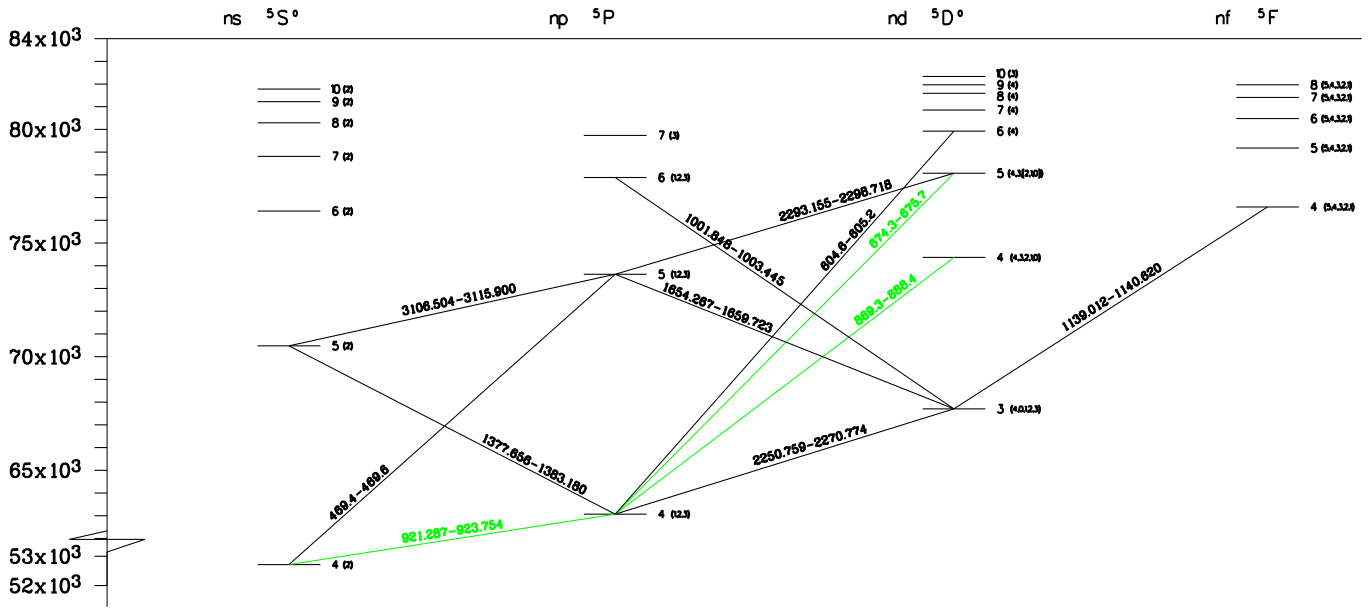,width=\hsize,height=8cm,angle=-90,clip=true}
\caption{ Grotrian diagram  of the Quintet system.
The decays that give rise to the lines of 
multiplets 1, 6, 8 are
indicated  in grey.
The partial Grotrian diagram of the quintet system, including the lines 
of astrophysical interest, was adapted from 
Atomic energy-level and Grotrian diagrams 
by \citet{bashkin}.
}
\label{grotdia}
\end{figure*}

\subsection{Choice of \ion{S}{i} lines}

In the \ion{S}{i} spectrum the lowest levels transitions belonging to the
triplet system lie in the UV below 200 nm and the ``raie ultime'' 
is at 180.731 nm; however, in this region
the flux of  F-G stars is too low  to be
easily observable.  
The S abundance in these stars can be derived from the strongest 
lines in the optical 
range, which belong to the quintet system with the lowest energy level at 
52623.640  cm$^{-1}$.

A preliminary selection of \ion{S}{i} lines at wavelengths  
shorter than 950 nm was made
on the basis of their intensity in the 
Revised Multiplet Tables (RMT) \citep{rmt}, 
the Utrecht Solar Atlas \citep{utrecht},
\citet[][LW, hereafter]{lambertwarner},
and
\citet[][BQZ, hereafter]{bqz}.
The synthetic spectrum of the Sun was 
computed and compared with the solar spectrum  of the  solar atlas
of \citet{kurucz_sol}
and used as a guideline.

The strongest lines in the visual-near IR spectrum are those of Mult. 1,
2, 6, and 8. 
The lines of Mult. 2  were discarded; the strongest line,
469.4113 nm, is weak and blended  with   \ion{Cr}{i} 469.4099 nm in the Sun.
Some of the lines of Mult. 6  were also  discarded on the basis of their
appearance in the solar spectrum. Line 
868.046 nm is  blended with a \ion{Si}{i} line with 
uncertain log gf, and  its intensity in the solar spectrum does not
agree with what is predicted by the computed spectrum. 
The lines at
867.065 nm (RMT=867.019 nm (J=1-0), 867.065 nm (J=1-1), 867.137 nm (J=1-2)) 
were discarded, because 
they are too weak to be measured in the  spectra of metal-poor stars. 
The lines of Mult. 7 were not 
considered, since 
their low intensity in the solar spectrum  makes
them useless for sulphur measurements in metal-poor stars. 
We point out that 
the  intensities given in RMT for these lines  are not coherent with
the line strengths observed in the Sun.

Other rather strong lines, but with higher 
excitation potentials, are those of: 

\begin{itemize}
\item
Mult. 10: the 604.1 nm line is  weak in the solar spectrum (LW)
and blended with \ion{Fe}{i} according to the Utrecht Solar Atlas. 
We note that the \ion{Fe}{i} line is weak
in  our computed spectrum.
Identification of the  604.5 nm line as \ion{S}{i} is   doubtful 
according to LW; according to the Utrecht Solar Atlas
it is affected by telluric lines. However,  at this wavelength no
telluric lines are detected in our spectra of
fast rotators.
The line at 605.2 nm  is not blended in the Sun and our
computed spectrum provides an excellent  
fit; however,  its low intensity would make it  measurable only in 
moderately metal-poor stars.
The latter  two lines were used by Chen et al. (2002), who
discarded the 604.5 nm anyway in their solar analysis, because they noted 
the presence of
an unidentified  blend in  the blue edge of the line. 
They claim, however, that the two  lines provide  consistent 
S abundances in their stars, which suggests that the blending
feature disappears at metallicities just below solar.

\item
Mult. 13:  all the lines are 
weak in the solar spectrum (LW); the strongest line, 903.588 nm, 
is blended with a  \ion{Cr}{ii} line. 
A further reason for discarding this multiplet is that
LW noted that they  give a smaller S abundance in the Sun.

\item
Mult. 21:  all lines are  easily measurable in the Sun, 
but are too weak to be measured at the
resolution of our  spectra. 

\item
Mult. 22: all lines are  blended with stronger 
lines  in the solar spectrum, and their gf values are uncertain.

\end{itemize}

The selected lines for our S abundance measurements 
are those  of
Mult. 1, 6, and 8 listed
in Tables \ref{trans_920}, \ref{trans_870}, and \ref{trans_670}
respectively.
Concerning  the lines of Mult. 8, we note the inconsistency 
of the EWs  in the Utrecht Solar Atlas (0.5 and 1.2 pm)
with those of  LW  
(1.7 and 1.2 pm) and 
also with 
the relative 
intensities of the 674.8 nm and 674.3 nm lines in the RMT  (8 and 6).

\bigskip

\begin{table}
\caption{Transitions and log gf of Mult. 1:
LW = \citet{lambertwarner}; Wiese = \citet{wiese};
BQZ = \citet{bqz} and RL = \citet{ryde}.}

\label{trans_920}
\begin{center}
{\scriptsize
\begin{tabular}{rrcrrlr}
\hline
\hline
\\
      & transition           &$\lambda$     & LW          &Wiese & BQZ  & RL      \\
      &                      &         (nm)&             &      &      &         \\
\hline
\hline
\\                                                                 
4s-4p & $^5$S$^o$$_2$--$^5$P$_1$ & 923.7538 & --0.01 (*)  & 0.04 & 0.10 & 0.04  \\
      & $^5$S$^o$$_2$--$^5$P$_2$ & 922.8093 & ~ 0.23      & 0.26 & 0.32 & 0.25  \\
      & $^5$S$^o$$_2$--$^5$P$_3$ & 921.2863 & ~ 0.38      & 0.42 & 0.47 & 0.43  \\
\\                                                                                          
\hline                                                                                      
\hline                                                                                      
\end{tabular}
}
\end{center}
(*) This is the value given by \citet{lambertwarner};
\citet{lambertluck}.
\citet{nissen}
use the value +0.01.\\
\end{table}

\begin{table}
\tabskip=0.5pt
\caption{Transitions and log gf of Mult. 6:
LW = \citet{lambertwarner}; W = \citet{wiese}; 
Fr = \citet{fran87};
BQZ=\citet{bqz}; C = \citet{chen02,chen03}.}

\label{trans_870}
\begin{center}
{\scriptsize
\begin{tabular}{llcrrrrr}
\hline
\hline
\\
      & transition         & $\lambda$      & LW     & W  &  Fr  & BQZ    & C \\
      &                    &           (nm) &        &    &      &        &   \\
\hline
\hline
\\
4p-4d & $^5$P$_3$--$^5$D$^o$$_3$ & 869.3931 & --0.56 & --0.51 & --0.74        & --0.85 & --0.52    \\
      & $^5$P$_3$--$^5$D$^o$$_4$ & 869.4626 &   0.03 &   0.08 & --0.21        & --0.26 &   0.05    \\
\\                                                                                          
\hline                                                                                      
\hline                                                                                      
\end{tabular}
\\
}
\end{center}
\end{table}

\bigskip

\begin{table}
\caption{Transitions and log gf of Mult. 8:
LW = \citet{lambertwarner}; Wiese = \citet{wiese};
BQZ = \citet{bqz} and Ecu = \citet{ecuvillon}.}
\label{trans_670}
\begin{center}
{\scriptsize
\begin{tabular}{rrcrrrr}
\hline
\hline
\\
 & transition               &$\lambda$      &  LW    & Wiese  &  BQZ   & Ecu \\
 &                          &          (nm) &        &        &        &     \\
\hline
\hline
\\                                                     
4p-5d & $^5$P$_1$--$^5$D$^o$$_0$ & 674.3440 & --0.85 & --1.27 & --1.20 & --1.27  \\
      & $^5$P$_1$--$^5$D$^o$$_1$ & 674.3531 & - -    & --0.92 & --0.85 & --0.92  \\
      & $^5$P$_1$--$^5$D$^o$$_2$ & 674.3640 &  --1.12& --1.03 & --0.95 & --0.93  \\
      & $^5$P$_2$--$^5$D$^o$$_1$ & 674.8573 & --1.48 & --1.39 & --1.32 & - -     \\
      & $^5$P$_2$--$^5$D$^o$$_2$ & 674.8682 & --0.48 & --0.80 & --0.73 & - -     \\
      & $^5$P$_2$--$^5$D$^o$$_3$ & 674.8837 &   - -  & --0.60 & --0.53 & - -      \\
      & $^5$P$_3$--$^5$D$^o$$_2$ & 675.6851 & --0.94 & --1.76 & --1.67 & - -     \\
      & $^5$P$_3$--$^5$D$^o$$_3$ & 675.7007 & - -    & --0.90 & --0.83 & --0.81  \\
      & $^5$P$_3$--$^5$D$^o$$_4$ & 675.7171 & --0.40 & --0.31 & --0.24 & --0.33  \\
\\                                                                                          
\hline    
\hline    
\end{tabular}                                                                                 
}
\end{center}
Chen et al. (2003) ignore the 
fine structure of the line and adopt
log gf = --0.70 for 674.36 nm and log gf = --0.31 for 675.717 nm.
\end{table}

\bigskip

\subsection{Atomic data}

The lines of Mult. 1 
were measured in stellar spectra only by 
\citet{nissen} and 
\citet{ryde}.
They are strong but difficult to measure 
due to the blending and to the presence of telluric lines.
Line 
921.2863 nm is blended with the weak \ion{Fe}{i} line 921.2970 nm;
 line
922.8093 nm is near the core of Paschen $\zeta$ (922.9017 nm),
and  line  
923.7538 nm is  on the far wing of 
the same Paschen line and near the \ion{Si}{i} 923.8037 nm
line.

The lines of Mult. 1 are almost ten times stronger than those of Mult. 6
so that they allow measurement of S abundance
for very metal poor stars. 
The 
oscillator strengths of the lines
of this multiplet  measured by
\citet{wiese}
agree with  those computed by LW
and also with those
computed by BQZ, 
as can be seen from Table 
\ref{trans_920}. \citet{ryde}
use the NIST 
database \citep{nist}.
\citet{nissen} use \citet{lambertluck},
which are the 
same as LW, except for the  923.7538 nm line.
We adopt the  
log gf values of the  NIST database, which 
holds the values of \citet{wiese},
which are experimental
but of D quality, corresponding to a possible
error up to 50 \%.
The same choice has been adopted for the other multiplets.

In literature most  S abundances 
are derived from the two strongest lines of Mult. 6. 
These are the only lines used by Clegg et al. (1981),
by Fran\c cois (1987 and 1988) and
by Israelian \& Rebolo (2001).

Part of the differences
in the sulphur abundances found in the
literature using the lines
of Mult. 6 may be tracked back to
the
log gf values adopted by the different authors;
in fact these lines display the
largest discrepancies 
among available 
log gf values.

Clegg et al (1981) used  the values of 
\citet{lambertwarner},
which  are also used by \citet{lambertluck}
for the solar
spectrum and by 
\citet{nissen}.
Fran\c cois (1987) 
derived ``solar'' oscillator strengths 
from the Moon spectrum
obtained with the same spectrograph 
he used for the   stellar spectra.
Israelian \& Rebolo (2001)
used  \citet{wiese} 
\footnote{they used the  values of
VALD database  which for \ion{S}{i}  contains the data in \citet{kurucz},
which, in turn, contains the 
Wiese et al. (1969)  measurements.}.
\citet{bqz} computed 
theoretical oscillator strengths
using both a Hartree-Fock relativistic code (HFR)
and the formalism implemented in the SUPERSTRUCTURE (SST) code,
and obtained a good agreement between the
two approaches. They recommend
HFR 
values because they are available for a larger set of lines.
According to these authors, their log gf should in general be more accurate 
than those of both LW and \citet{wiese}. 
It is significant, however, that the two lines 869.40 nm and 869.47 nm are 
not retained in the BQZ determination of the solar S abundance.  
For this reason we decided not to adopt the oscillator
strengths of BQZ.

The lines of Mult. 8 are free of blends and arise from the same lower
level as those of Mult. 6,  so that the dependence of these two sets of lines
on  \teff and log g of the stars is the same. 
The oscillator strengths used by different authors are in agreement: 
LW,
\citet{wiese} and
BQZ, and also \citet{ecuvillon} who
adopted VALD data modifying them to obtain a good fit to the solar spectrum.
For Mult. 8 the log gf measured by \citet{wiese} 
and those computed by BQZ 
and LW are similar. The problem of the high discrepancies among log gf 
values appears to be restricted to the two lines of Mult. 6.
Our tests  show that the BQZ log gf 
of Mult. 6 lines  produces S abundances 
not coherent with those derived from Mult. 8, strengthening our 
choice of not
adopting this source.

From Tables 2, 3, and 4 we conclude  that literature
data need  not be scaled, because the adopted 
log gf  values are compatible.
The differences in adopted log gf values
are  within the random errors of S abundance determinations.

\subsection {S abundance in the Sun}

We note small 
differences in the solar S abundance used by the quoted authors.
The value A(S)$_\odot=7.21$ 
(Anders \& Grevesse 1984 and 1989) 
is used by  \citet{fran87,fran88} and
\citet{israelian}.
Chen determined A(S)$_\odot$=7.20,
a value adopted by 
\citet{nissen} and  
\citet{ryde}; 
\citet{takeda} use A(S)$_\odot$=7.21 for their HIRES sample, 
and 7.22 for their OAO sample because this is the sulphur solar
abundance they derived.
The only discrepant value is BQZ A(S)=7.33 
adopted by \citet{grevesse96} and \citet{grevesse98}.

\citet{chen02} 
in the Notes of Table 4 say that ``the $\alpha$ enhancement 
for S is calculated assuming A(S) = 7.33 for the Sun
following \citet{grevesse98}''.
In his models Kurucz retains A(S)=7.21, i.e. the value
of \cite{lambertwarner}
and \citet{anders};
the same selection is made by \citet{lodders} for the solar photospheric value.
We finally recall A(S)=7.14 by \citet{asplund} who 
used a 3D hydrodynamical model.
We did not consider this value because it is not directly
comparable to our results obtained from 
1D models.

\subsection{Predicted intensity of \ion{S}{i} lines in stellar spectra}

In the metallicity range of our sample,
the use either of 
$\alpha$-enhanced or  non  $\alpha$-enhanced models 
affects the derived
sulphur abundances by a few hundredths of dex
at most, as already shown by  \citet{chen02}.

Sulphur abundance is dependent on \teff and log g,
as well as on the metallicity
of the star. In fact A(S) increases with increasing \teff, while it
decreases, at constant \teff value, with decreasing gravity.

\section {Analysis}

\subsection{Model atmospheres and synthetic spectra}

For each star we computed a model atmosphere using 
version 9 of the ATLAS code  \citep{kurucz} running under
Linux  \citep{sbordone}. We used the updated Opacity 
Distribution Functions of \citet{c_k} with
microturbulent velocity of 1 \kms and enhancement of $\alpha$
elements.
The synthetic spectra were computed using the SYNTHE suite
\citep{kurucz} running under Linux \citep{sbordone}.

\subsection{Line profile  fitting}

In the  present investigation we decided to determine
abundances by making use of line profile fitting. This approach is
required  for the lines of Mult. 1 which, as 
discussed above, are affected by the wings of Paschen $\zeta$;
and it is also desirable for the  lines of Mult. 8, which are affected by
fine structure  splitting, although one could ignore
the fine structure since the lines are weak.
In principle, the   lines of Mult. 6 which were used are isolated 
and could be treated efficiently by simply measuring the
equivalent width. However, we decided to also use line profile fitting
in this case for  two  reasons: first
for homogeneity with what was done for the other lines, and second 
because these lines are often weak, especially
at the lower metallicities. The inclusion of fitting
continuum and neighbouring lines with a synthetic spectrum 
greatly improved the stability of the fitting procedure with respect to
fitting a single gaussian to measure the equivalent width.

We developed a line fitting code which performs a $\chi^2$ fit
to the observed spectrum. The best fitting spectrum is 
obtained by linear interpolation between three synthetic spectra
which differ only in their S abundance; the minimum $\chi^2$
is sought numerically by making use of the {\tt MINUIT} routine
\citep{minuit}. As pointed out by \citet{BC03} the 
$\chi^2$ theorems do not apply to the fitting of spectra,
because the pixels are correlated\footnote{In fact, on the one hand,
in most spectrographs the slit projects on at least two pixels, thus
making the signal in neighbouring pixels correlated; on the other
hand, one usually works on spectra which were rebinned to a
constant wavelength step, in order to allow the coaddition of different
spectra, often with an oversampling factor of at least 2. Therefore
neighbouring pixels are strongly correlated. }.
However, $\chi^2$ minimization may still be used
to estimate parameters (in our case abundances), but 
errors cannot be deduced from $\chi^2$ \citep{numrec}.
In fact, as described in the next section, we resort
to Monte Carlo simulations to estimate errors.

\begin{figure}
\psfig{figure=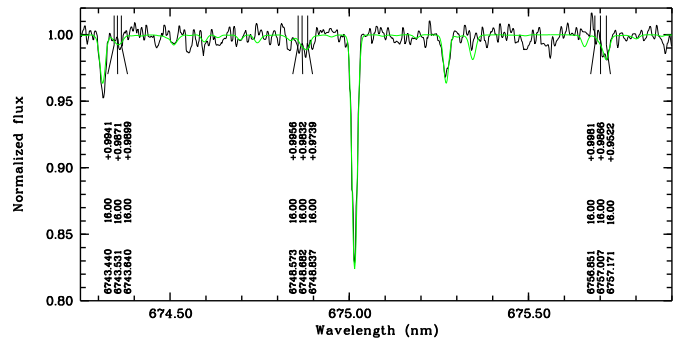,width=\hsize,clip=true}
\caption{ Fit in the region of 670 nm of the star \object{HD 25704}
(black observed spectrum, grey fitted spectrum).
The \ion{Fe}{i} 675.0152 is in good agreement with the [Fe/H] utilized, 
so we fitted the
whole range including the \ion{S}{i} 
lines of Mult. 8 and the lines of other elements
as well.
}
\label{S6700}
\end{figure}

\begin{figure}
\psfig{figure=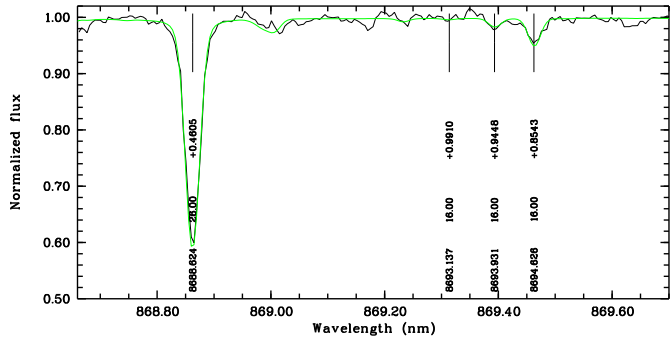,width=\hsize,clip=true}
\caption{ Fit in the region of Mult. 8 of the star \object{HD 10607};
black is the observed spectrum and grey the fit.
}
\label{S8700}
\end{figure}

The fitting code works interactively under {\tt MIDAS}: the user
selects the spectral region to be fitted by using the graphic cursor.
For 
sulphur we selected regions 
containing one or more \ion{S}{i}  lines and in some
cases also  lines of
other elements. 
In the  region of Mult. 8, all the 3 \ion{S}{i}  lines\footnote{
At our resolution, the fine structure is not resolved,
although the ``lines'' are really a blend of
three lines.} were fitted simultaneously. 
This region also includes  
the  \ion{Fe}{i}  675.0152 nm line and some weak lines
of other elements. 
An example  is given in Fig. \ref{S6700}.
The presence of the \ion{Fe}{i}  line
does not disturb the fit; on the contrary, it helps
make it more stable, especially 
in cases in which 
the adopted 
Fe abundance fits the observed line  well.
In those cases in which  the adopted Fe abundance is not
in agreement with the line strength of this
particular line, or  when only  
one of the \ion{S}{i}  lines is detectable or the spectrum
is affected  by bad pixels, cosmic ray hits, etc.,
the fit was made on a single
\ion{S}{i} line.
The same criterion was used for the \ion{Fe}{i} 868.8624 nm:
for the lines of Mult. 6 an example is shown in Fig.
\ref{S8700}.

\begin{figure}
\psfig{figure=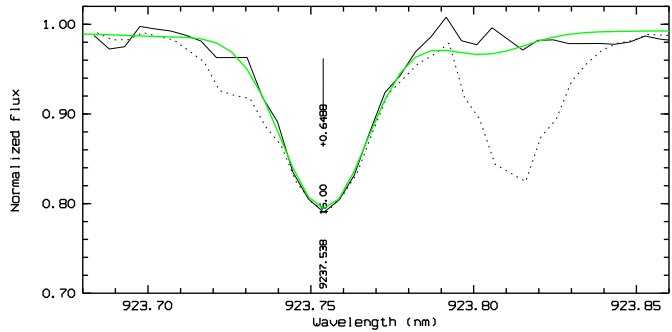,width=\hsize,clip=true}
\caption{ Fit in the region of Mult. 6 of the star \object{HD 10607}.
The solid line is the observed spectrum after subtraction of telluric lines;
dotted line  is observed spectrum,
grey is the fit.}
\label{S9200}
\end{figure}

In the region of Mult. 1, each \ion{S}{i} 
line was fitted individually because of the
telluric lines affecting the spectrum.
In a first step the regions affected by 
telluric lines were identified
by using the spectrum of a fast rotator, and if
a telluric line was blending an \ion{S}{i} line,
this was not used.
An example of a fit of a Mult. 1 line is shown in Fig. \ref{S9200}.

In a second step we tried to use
the spectrum of a fast  rotator, suitably scaled,
to remove the telluric absorptions which are affecting \ion{S}{i} 
lines.
The spectra of fast rotators at our disposal 
were not ideal since they were seldom
observed on the same night as our programme  stars
and hardly ever at the same airmass.
Under these conditions  the removal of the
tellurics is rather
unstable.
In fact, it can happen that their removal
produces a clean
\ion{S}{i} line, but in some cases this line implied an
abundance in dire disagreement with that measured from
lines unaffected by the tellurics, and thus
the procedure was unsuccessful. 
We conclude that,  under these conditions,
we would retain the measures  
on lines from which a contaminating telluric was removed
only in those cases in which at least one of the three lines
was unaffected by tellurics and the abundances derived from 
both contaminated and uncontaminated lines
were in agreement.

\begin{figure}
\psfig{figure=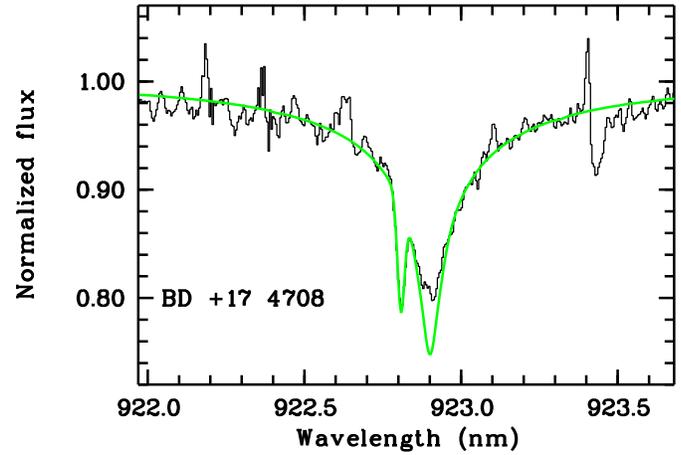,width=\hsize,clip=true}
\caption{Fit of the 922.8nm \ion{S}{i} line in star
\object{BD +17 408}. It may be appreciated that
the synthetic spectrum provides a good fit to the
wings of Paschen $\zeta$, while not to the core.
This is not surprising since our synthetic
spectra are computed
in LTE and the core of Paschen lines is predicted
to be affected by NLTE effects (see for
example Johnson \& Kinglesmith 1965).
}
\label{Pz}
\end{figure}

In theory one should be able to 
deduce the sulphur abundance for lines of Mult. 1 
from the equivalent widths (EWs).
In practice,  the three sulphur lines 
lie next to the Paschen $\zeta$  line.
The 922.8093 nm line is next to the core of
Paschen $\zeta$ and 
it is impossible to measure its true equivalent width.
But the 921.2863 nm and 923.7538 nm \ion{S}{i} lines are
also affected by the  
wings of Paschen $\zeta$ and, moreover, the former 
is blended with
a weak \ion{Fe}{i} line (921.297 nm), and
the latter is blended  
with  \ion{Si}{i} 923.8037 nm.
These two blending lines are so weak
that they could be ignored.

\begin{figure}
\psfig{figure=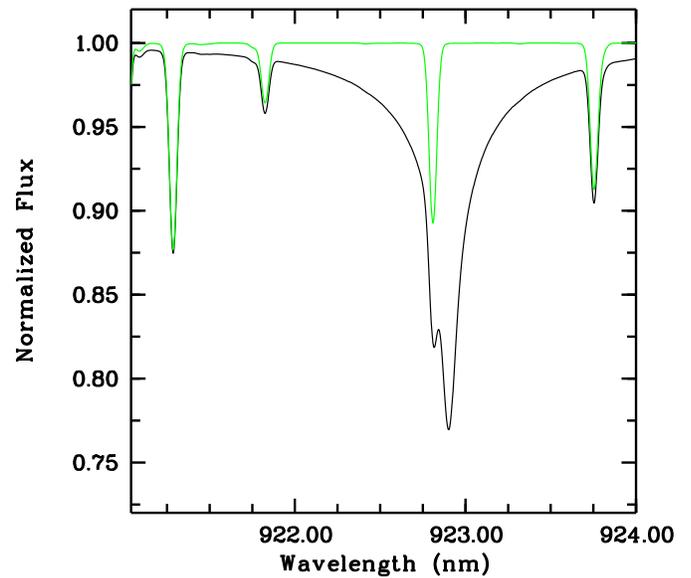,width=8.8cm,clip=true}
\caption{ Two spectra with parameter: Teff=6016 K,
log g=4.04, and [Fe/H]=--1.5, the G 126 --62
star parameters. The black one is the one we used to
fit the sulphur abundance, while the grey one is built without
the 922.9017 nm Paschen $\zeta$.}
\label{h_noh}
\end{figure}

To illustrate the effect of Paschen $\zeta$ and support
our statement that spectrum synthesis is needed
to study sulphur in this region, 
we computed two synthetic spectra with Teff=6016  K, log g=4.04,
and [Fe/H]=--1.5 (parameters of star \object{G 126--62}), one with 
Paschen $\zeta$
and
the other without it
(see Fig. \ref{h_noh}).
It may be appreciated 
that  the EW of the 921.2863 nm line is almost the same
in both cases, while the
EW of the  923.7538 nm is larger in the spectrum computed 
without the hydrogen line. 
Going from EWs to abundances, one finds
that matching a measured EW to one computed
ignoring Paschen $\zeta$, an S abundance
that is lower by 0.08 dex is derived.

In measuring the EW, one is forced to use the wing
of Paschen $\zeta$ as a local continuum, effectively
underestimating the true EW.
We performed a few experiments by measuring this
EW on the synthetic spectrum and found that 
this underestimated the S abundance
by an additional 0.08 dex.
This effect of Paschen $\zeta$ on
the 921.2863 nm sulphur line is instead negligible,
as already stated by \citet{nissen}.

With this discussion we hope we have convinced
the reader that the use of spectrum synthesis
is definitely preferable to the use of EWs 
to derive S abundances from  lines of Mult. 1.
Table \ref{sdat} reports
sulphur abundances derived (when measurable) from each multiplet,
the weighted average and the relative error.

\section {Errors and sensitivity of abundances to stellar parameters}

For reasons stated in the previous section,
one cannot rely on the $\chi^2$ theorems to obtain 
error estimates; we, therefore,
resorted to a 
Monte Carlo method.
We first estimated the
statistical error:
a synthetic spectrum with parameters 
representative of our stars
was computed 
and  Poisson noise was injected
to obtain the desired S/N ratio. We shall
refer to this as  a ``simulated observation''.
This spectrum was fitted in a similar way to the
observed spectrum and the
fitted parameters were derived.
The process was repeated 500  times for the lines
of Mult. 8, which are the weakest of
the three multiplets considered here.
The results are given in Table \ref{tabsigma}, 
where the mean derived abundance and
the dispersion around this mean value
are reported for different  S/N ratios.
The dispersions may be taken as estimates
of the statistical errors.
It is interesting to point out that 
small offsets exist in the mean
derived abundance: at low S/N ratios
the S abundance is overestimated by hundredths of dex,
while at very high S/N ratios
the abundance is underestimated by 0.01 dex.

\begin{table}
\caption{Spectrum with T=5800 K, log g=4.25, [Fe/H]=--1.5, [S/H]=-1.10
for the lines of Mult. 8. Each
Monte Carlo simulation includes 500 events.}
\label{tabsigma}
\begin{center}
\begin{tabular}{rll}
\hline
\hline
\\
S/N &  A(S)  & error \\
\hline
\hline
\\
 300 & 6.11  & 0.05\\ 
 250 & 6.10  & 0.06\\ 
 200 & 6.10  & 0.07\\
 150 & 6.10  & 0.08\\
 100 & 6.10  & 0.13\\
  80 & 6.10  & 0.17\\
  50 & 6.14  & 0.21\\ 
  30 & 6.18  & 0.29\\
\\
\hline
\hline
\end{tabular}
\\
\end{center}
\end{table}

In
the case of incorrect atmospheric parameters,
we also tried to estimate
systematic errors, or rather
the combination of systematic
and statistical errors.
The same procedure as described above was applied.
However, we tried to fit our
simulated observations 
with synthetic spectra whose atmospheric
parameters were different from those of the
simulated observation. We explored the effect
of one parameter at the time:
temperature, gravity, and Fe abundance.
The systematic errors are estimated as
the  difference between
the  sulphur abundance 
of the simulated spectra and the mean fitted 
sulphur abundance. 
The results for the lines of Mult. 8 are given in Table \ref{sistemerr}.
We further checked the variation of the systematic error with \teff
when changing \teff in the range 5000 to 6000 K by simply using
the WIDTH code. An error of 100 K implies an error in the
sulfur abundance of
$\sim 0.05$ dex around 6000 K and $\sim 0.09$ dex around 5000 K.
These results, together with those of Tables 
\ref{tabsigma} and
\ref{sistemerr}, have been used to estimate the errors on our
S abundances, as described in the next section. 

\begin{figure}
\psfig{figure=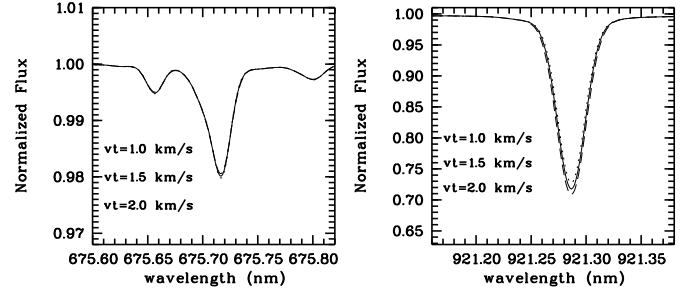,width=\hsize,angle=0,clip=true}
\caption{ Microturbulence effect on sulphur lines; the synthetic spectra
plotted
have: Teff=5810K, log g=4.50, [Fe/H]=--1.0, which are the parameters of
the star \object{HD 205650}. This star has been selected to show microturbulence
effect, because its parameters fall near the average of
our sample.}
\label{vturb}
\end{figure}

\begin{table}
\caption{Spectrum with T=5800 K, log g=4.25, [Fe/H]=--1.5, [S/H]=--1.10. Each
Monte Carlo simulation has 500 events}
\label{sistemerr}
\begin{center}
\begin{tabular}{lllllll}
\hline
\hline
\\
T    & log g & [Fe/H] & $\xi$ & A(S) & $\sigma_{ran}$ & $\sigma_{sys}$ \\
K    & cgs\\
\hline
\hline
\\
5600 & 4.25 & --1.5 & 1.00 & 6.28 & 0.10 & +0.17 \\
5700 & 4.25 & --1.5 & 1.00 & 6.17 & 0.10 & +0.06 \\
5900 & 4.25 & --1.5 & 1.00 & 6.05 & 0.11 & --0.06\\
6000 & 4.25 & --1.5 & 1.00 & 6.00 & 0.11 & --0.11\\
\\                                                 
5800 & 4.00 & --1.5 & 1.00 & 6.03 & 0.10 & --0.08\\
5800 & 4.50 & --1.5 & 1.00 & 6.18 & 0.10 & +0.07 \\
\\                                                 
5800 & 4.25 & --1.0 & 1.00 & 6.34 & 0.10 & +0.23 \\
5800 & 4.25 & --1.3 & 1.00 & 6.14 & 0.08 & +0.03 \\
5800 & 4.25 & --1.4 & 1.00 & 6.12 & 0.09 & +0.01 \\
5800 & 4.25 & --1.6 & 1.00 & 6.10 & 0.09 & --0.02\\
5800 & 4.25 & --1.7 & 1.00 & 6.08 & 0.10 & --0.03\\
5800 & 4.25 & --2.0 & 1.00 & 6.01 & 0.14 & --0.10\\
\\                                                 
5800 & 4.25 & --1.5 & 1.50 & 6.18 & 0.14 & +0.07 \\
\\
\hline
\hline
\end{tabular}
\\
\end{center}
\end{table}

Another 
systematic error is due to the  microturbulent velocity.
However, this is non-negligible only for the lines of Mult. 1, as
the lines of Mult. 6 and 8 are weak 
and may be considered insensitive
to  microturbulence for all practical purposes.
In fact, for a change of 0.5 \kms, the Monte Carlo
simulation  of 500 events shows a systematic error that is smaller than
the random one.

In order to test the sensitivity of 
the lines of Mult. 1 to microturbulence,
we performed a  Monte Carlo simulation of 500 events and S/N=150.
The parameters 
of simulated observations were \teff=5800 K, log g=4.25, 
[Fe/H]=--1.5 and log (S/H) + 12 = 6.11, and
$\xi = 1.0$ \kms.
A run using synthetic spectra with the same
parameters as the simulated observation (except for S abundance)
provides
A(S) =  log (S/H)+12$=6.11\pm 0.08$. 
If, in the same conditions, we use 
synthetic spectra with microturbulence of 1.5 \kms instead, 
we obtain A(S) $=6.05\pm 0.03$. 
Therefore, an error of 0.5 \kms in the microturbulence
results in an error of 0.06 dex in the S abundance.
This systematic error is comparable
to the random error.
In Fig.\ref{vturb} the variation of the residual intensity 
for a change of 0.5 \kms is shown.

A further source of systematic errors
may  be NLTE effects in these lines. 
\citet{takeda} studied the NLTE effects for
the lines of Mult. 6 and concluded
that they are
negligible.
\citet{chen02} claim that
 the same must be true for the weaker lines of Mult. 8, 
which have the same low EP. 
\citet{nissen} claim that 
NLTE should   be small for the lines of Mult. 1
since, in their analysis, the
S abundances derived from these
lines agree with those derived from the lines of
Mult. 6. 
The NLTE computations of \citet{takeda} allow us
to neglect NLTE effects with confidence 
for the lines of Mult. 6, the arguments of
\citet{chen02} and \citet{nissen} justify
neglecting them for the lines of Mult. 1 and Mult. 8,
although in these cases detailed NLTE computations
would also be desirable. In particular, from our own analysis,
the lines of Mult. 1 seem to provide S abundances which
are about 0.17 dex higher than those derived from Mult. 8 and 
Mult. 6; this falls within the observational errors and may 
not be significant at all. However, it would be interesting
to investigate possible NLTE effects.
On this point
see also the discussion in \citet{ryde}.

Finally, one should take the possible effects
of granulation \citep{asplund} into account. Such
computations for these lines are not available at present, and we shall
therefore ignore them, although we are aware that they
might be relevant.

\section{Sulphur abundances}

The lines of Mult. 8 are not detectable  for the most metal--poor
stars in our sample. 
They are detected only in  28 out of 74 stars with a 
spectrum covering this spectral region, and  
the most metal-poor of them has [Fe/H]=--1.67.

The stars for which we had a spectrum 
covering the lines of  Mult. 6 
were fewer, i.e. only the 50 stars with UVES spectra
in common with \citet{gratton}.
We derived sulphur abundances using
the lines of Mult. 6 for 21 of them, the most metal-poor 
with  [Fe/H]=--1.84.

Spectra covering the line range
of Mult. 1 were available
for 36 stars \footnote{a slightly different setting was used
in the different runs of programme 165.L-0263 and the
setting used in the first run did not cover the region
of the \ion{S}{i} Mult. 1 lines.}.
We detected these lines  in 28 stars. 
In the remaining 8 stars  we did not measure sulphur from
the lines of Mult. 1
because the telluric lines blend  all three
lines of the multiplet, so 
their removal is unsatisfactory.
The most metal-poor star in our sample has
[Fe/H]=--2.43.

\begin{figure}
\psfig{figure=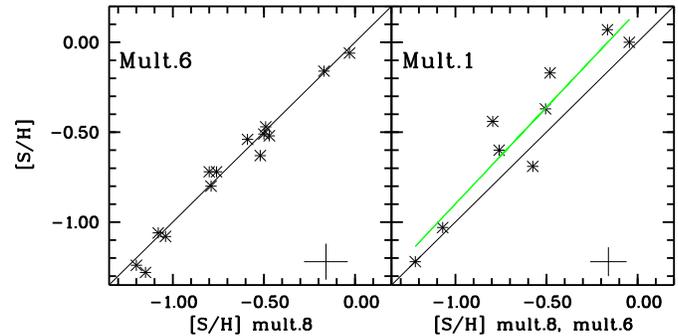,width=\hsize,clip=true}
\caption{
Left:
[S/H] derived from the lines of Mult. 6 versus
[S/H] derived from the lines of Mult. 8 for all the
stars with measurements in both multiplets. 
Right:
[S/H] derived from the lines of Mult. 1 versus 
mean [S/H] derived from the lines of Mult. 8 and Mult. 6 for all
the stars with measurements in all three multiplets. 
In both plots the solid line is the bisector;
the cross at bottom is a representative error bar.
The bisector shows the good agreement of [S/H] between
Mult. 8 and Mult. 6 (left). The systematic difference of
[S/H] derived from Mult. 1 is shown by the linear
correlation (grey line).
}
\label{s870vss670}
\end{figure}

The concordance  of the sulphur abundances deduced from 
the lines of Mult. 6 and Mult. 8.
is generally good. There are 14 stars in which the lines
of both multiplets were measured and the
mean difference in [S/H] is $0.014\pm 0.06$ (see Fig. \ref{s870vss670}).

\begin{figure}
\psfig{figure=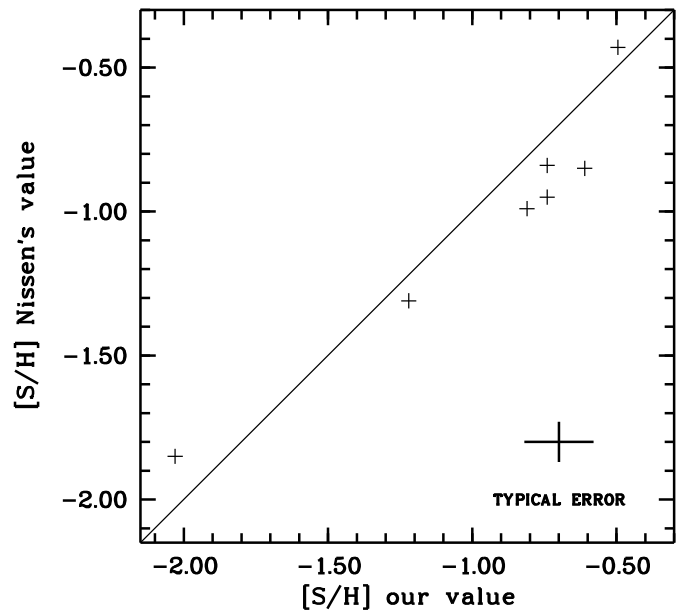,width=8.8cm,clip=true}
\caption{ Comparison of [S/H] value from our measurements and
the  determination of \citet{nissen}.
}
\label{nisnoiconf}
\end{figure}

\begin{figure*}
\psfig{figure=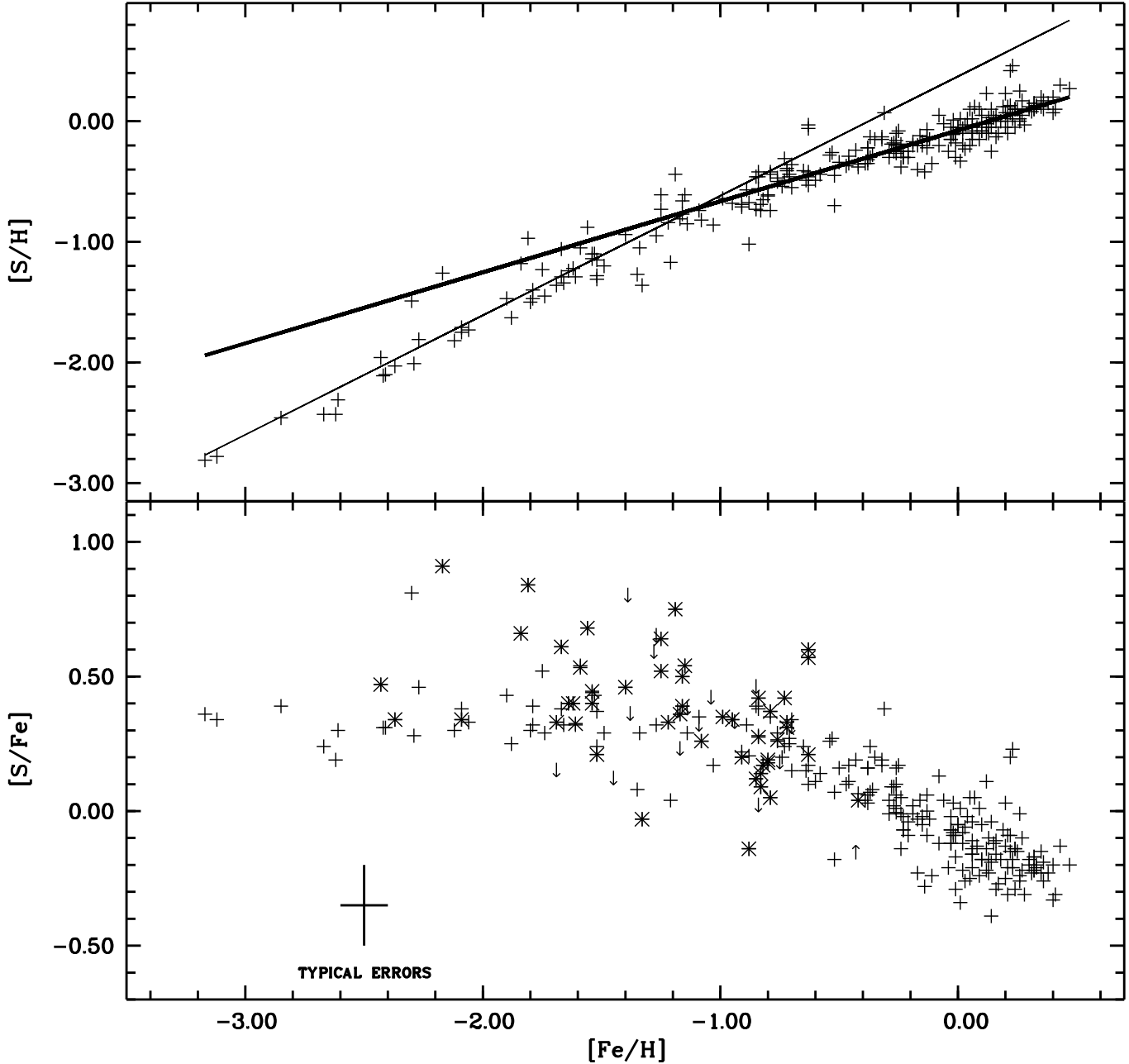,width=\hsize,clip=true}
\caption{ Bottom panel: [S/Fe] versus [Fe/H]. The measures of the present
paper are indicated as asterisks, crosses are the
data taken from the literature (Table \ref{sdat}).  
The typical error bar is shown in the lower left corner.
Top panel: [S/H] versus [Fe/H] for all stars considered.
Two different linear trends can be distinguished:
the thick  line is a  fit to all stars with
[Fe/H]$> -1$, the thin line to those with
[Fe/H]$\le -1.0$.
}
\label{sfeall}
\end{figure*}

\begin{figure}
\psfig{figure=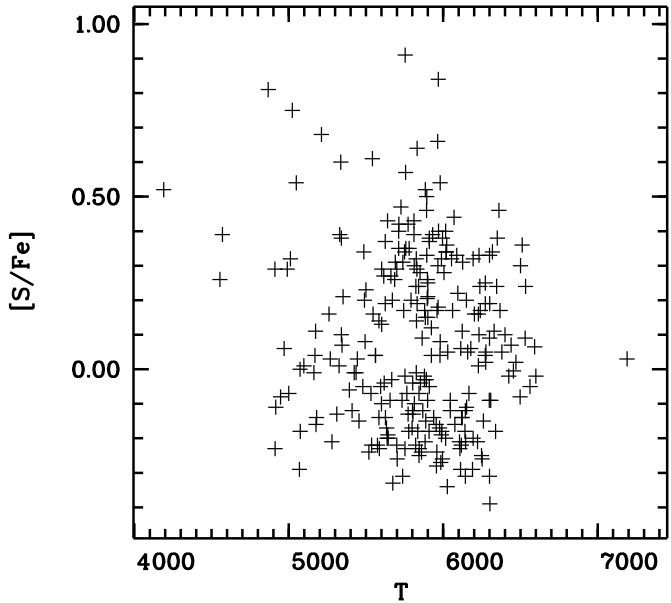,width=\hsize,clip=true}
\caption{The [S/Fe] ratio versus effective temperature
for all the stars in Table \ref{kinetab}. No trend
is discernible.}
\label{tvsfe}
\end{figure}

\begin{figure}
\psfig{figure=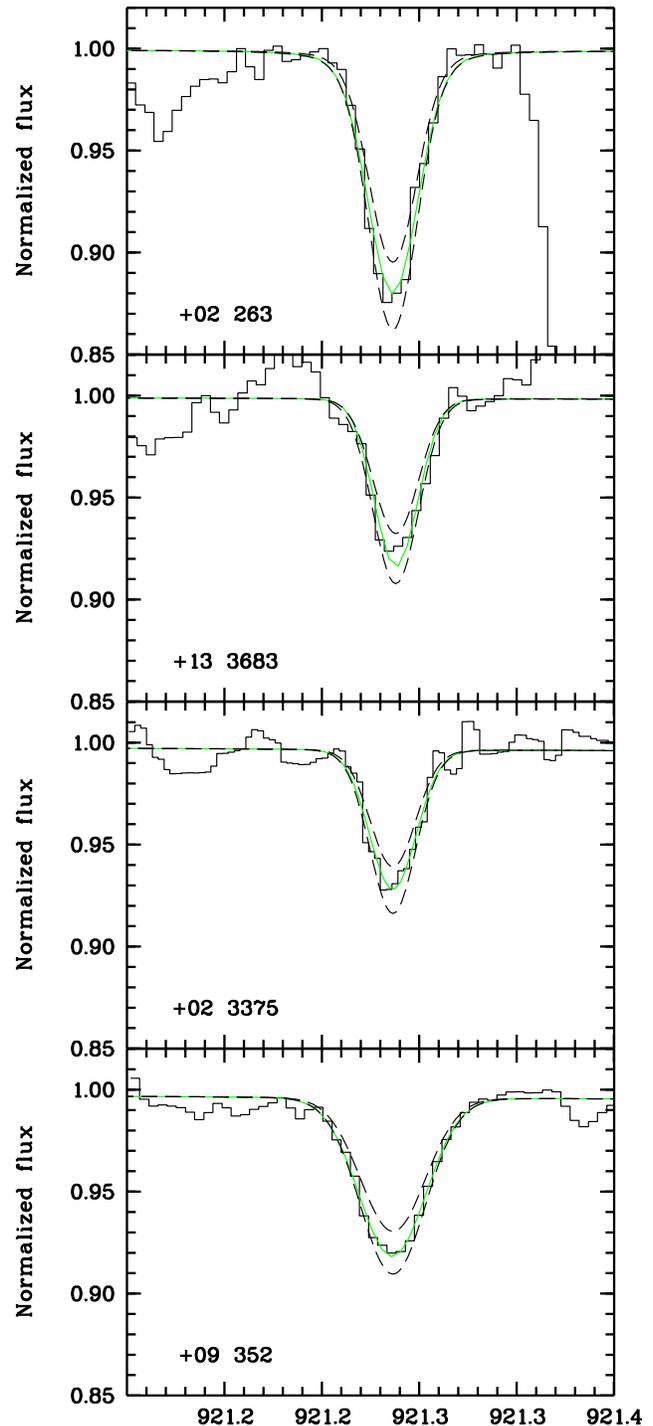,width=0.95\hsize,clip=true}
\caption{ 
Fit of the 921.2 nm line (grey line) of four of the most metal poor
stars that lie in different places in the plot [S/Fe] versus
[Fe/H]. 
Dashed lines are synthetic spectra
computed with $\pm 0.1 $ dex the best fitting abundance.
As one can see the four fits are good, and the abundance
deduced by these fits should be reliable.}
\label{badfit}
\end{figure}

\begin{figure}
\resizebox{\hsize}{!}{\includegraphics[clip=true]{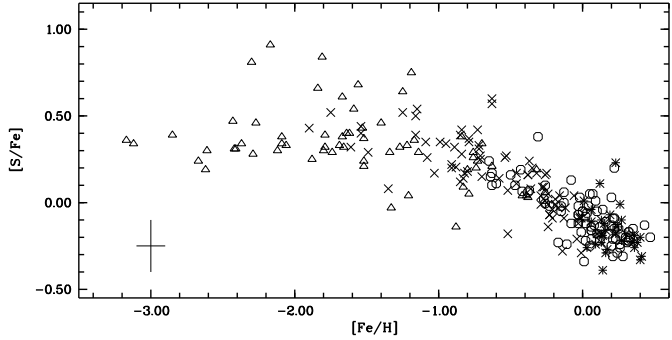}}
\caption{[S/Fe] versus [Fe/H] for the stars classified
on the basis of their Galactic orbits: open circles are
the thin disc stars, crosses the dissipative
component, triangles the accretion component,
and the asterisks are the stars which do not
fall in any of these categories.}
\label{sfe_pops}
\end{figure}

\begin{figure}
\resizebox{\hsize}{!}{\includegraphics[clip=true]{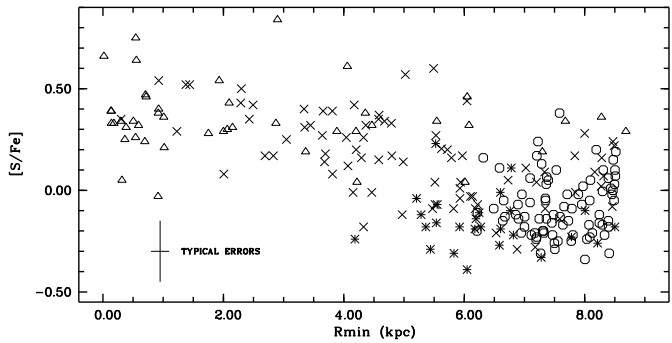}}
\caption{[S/Fe] as a function of perigalactic distance Rmin, in kpc.
The different populations are marked as in Fig. \ref{sfe_pops}.}
\label{rp_sfe}
\end{figure}

\begin{figure}
\resizebox{\hsize}{!}{\includegraphics[clip=true]{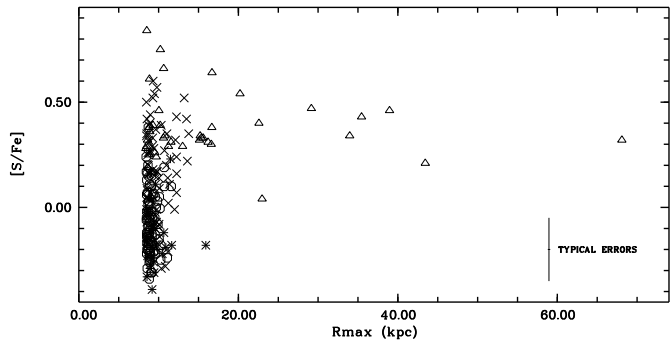}}
\caption{[S/Fe] as a function of apogalactic distance Rmax, in kpc.
The different populations are distinguished as in Fig. \ref{sfe_pops}.}
\label{ra_sfe}
\end{figure}

\begin{figure}
\psfig{figure=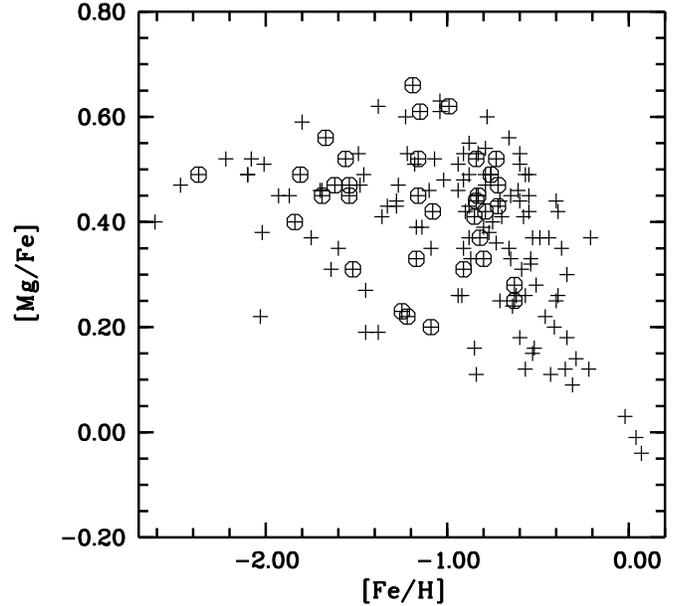,angle=0,clip=true}
\caption{[Mg/Fe]  versus [Fe/H] from 
\citet{gratton}; circled stars are those in common with the present
study. }
\label{mgfe_gratton}
\end{figure}

\begin{figure}
\psfig{figure=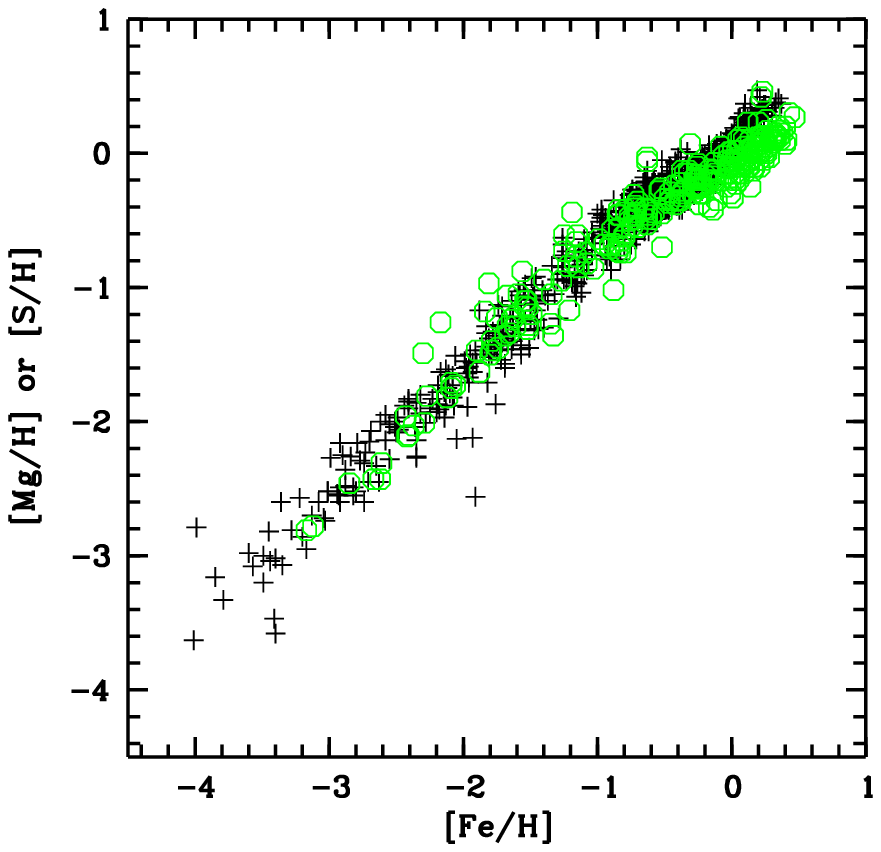,width=\hsize,clip=true}
\caption{ [Mg/H] versus [Fe/H] for the stars 
in the compilation of \citet{venn04} (crosses)
and [S/H] versus [Fe/H] (open hexagons) from the compilation
in Table \ref{kinetab}.
}
\label{mgvsfe}
\end{figure}

No star has sulphur detection  only in Mult. 8 and in Mult. 1.
Five  stars have sulphur detections in Mult. 6 and Mult. 8.
Nine stars have sulphur detection in all  three multiplets. 
For these 9 stars,
we plot  the mean [S/H] measured
from Mult. 6 and Mult. 8 versus the value measured from 
Mult. 1 in Fig. \ref{s870vss670}.
A simple regression provides a slope of 1.07 ($\pm 0.15$) 
and an offset of 0.18 ($\pm 0.11$) dex.
Due to the size of the errors, it is not clear whether this
offset is significant. 
We tested this effect by applying it to  
all the measurements from Mult. 1, but
none of our conclusions is affected by it. We therefore
decided not to apply this small and uncertain correction
to our measurements. 
The two stars which show the largest discrepancy,
greater than 0.3 dex, are \object{HD 17072} and \object{HD 204155}.
\object{HD 17072} is the only giant star in our sample with
a microturbulent velocity of 2.1 \kms \citep{carney}.
\object{HD 204155} is a dwarf with a microturbulent 
velocity of 0.98 \kms \citep{gratton}.
For both stars the S abundance deduced from the
lines of multiplets 6 and 8 are in excellent agreement. 
In both cases an increase of the microturbulent
velocity on the order of 1 \kms would bring
the S abundance deduced from the lines of Mult. 1
in agreement with that deduced from the other lines,
but we have no justification for such an increase.

The last but one column in Table \ref{sdat} provides the
final adopted
value of [S/Fe] for each star, which
is simply the mean of the measurements we made,
and the last one is the associated error.
There are 
50 stars with an [S/H] determination.
We could not determine sulphur abundances for all the stars in the sample,
but we give some upper limits. For some stars, no determination or
upper limit was possible either because of bad pixels, or because the
signal-to-noise ratio was too low. For one star (\object{HD 83220}),\ 
we give a lower limit because
a cosmic ray lies on the core of the sulphur line (675.7 nm), which
is clearly present.
In those cases for which upper limits were available
for some lines and detections for others, we ignored
the upper limits.

To compute an  error estimate
for  our sulphur measurements,
we used the S/N ratios given in Table \ref{data},
and used the results of the Monte Carlo simulations in Table
\ref{tabsigma} to estimate the random error
which was added linearly to the systematic errors due
to \teff and log~g errors, as discussed in the previous section.
Finally, the result was divided by $\sqrt{n}$,
where $n$ is the number of multiplets for which
we have an S measurement.
Errors on upper limits were
computed  in the same way, but only for the stars
for  which we have no measures.
Our error estimates for [S/H] are provided in the last column
of Table \ref{sdat}.

\subsection{Comparison with other authors}

We share few stars 
with other authors:
one star (\object{HD 194598}) is in common with \citet{israelian} and 
[S/H] is in
good agreement (difference = 0.04 dex); 
one star (\object{HD 22879}) is in common with \citet{chen03} with
[S/H] in good
agreement (difference of 0.01 dex);
7 stars are in common with \citet{nissen},
the mean difference is
$+0.08\pm 0.15$, with  no evidence of any systematic effect
(see figure \ref{nisnoiconf}).

\section{Results}

Our sample is probably
too small to adequately investigate the evolution
of sulphur abundances.
It is, therefore, worthwhile to combine
our sample with the other measures available in the literature.
Since there is little overlap among the different
samples in the literature,
it is possible to gain considerable insight
by combining them.
We took the data from
\citet{israelian},  \citet{takeda},
\citet{chen02}, \citet{chen03}, \citet{ryde},
\citet{nissen}, and
\citet{ecuvillon} and combined them with our own.
We considered the log gf values used by the different
authors and concluded that all the above S measurements
are on the same scale, since the differences in adopted
log gf are usually smaller than the abundance errors 
(see Sect. 4.2).
When a star was observed by several authors, we proceeded
as follows: if the star was also in our own sample
we kept our measurements; if different authors
adopted similar atmospheric parameters we
took a straight average of the different measurements;
if, instead, the adopted atmospheric parameters
were significantly different, we adopted only one
of the measurements according to our perception
of which was the most reliable.
For example, for the stars common to 
\citet{chen02} and \citet{nissen}, we preferred 
\citet{nissen}, since most likely VLT data are of
higher quality, and so on. This procedure
is somewhat arbitrary, but is preferable to 
taking a straight average of S abundances
derived by adopting different effective temperatures.
Moreover, this regards a very limited number
of stars, and the general conclusions we derive
do not depend on these choices. 
In this way we assembled 
a sample of 253 stars with a unique
[S/H] value, and this compilation
is given in Table \ref{kinetab}.
The overlap among
the different investigations is minimal:
only  29  stars are
analyzed by more than one author.
In column 11 of  Table \ref{kinetab} we provide the reference
to all the papers which provide S measurement for a given star.

The data of \citet{israelian}  and  \citet{ryde}
are not in agreement,
even if they use the same \teff. We adopted the mean value.
The two authors considered two 
different regions: \citet{israelian} used  the lines of Mult. 6, 
while \citet{ryde} used the lines of Mult. 1. 
We note that sulphur abundances in \citet{ryde} 
are systematically lower than those in \citet{israelian}.

\object{HD 9826} presents a different sulphur abundance in \citet{chen02} (+0.16) 
and in \citet{ecuvillon} (--0.20), even if the same lines are 
considered in the two papers.
\object{HD 217107} presents the same problem.

\begin{figure*}
\psfig{figure=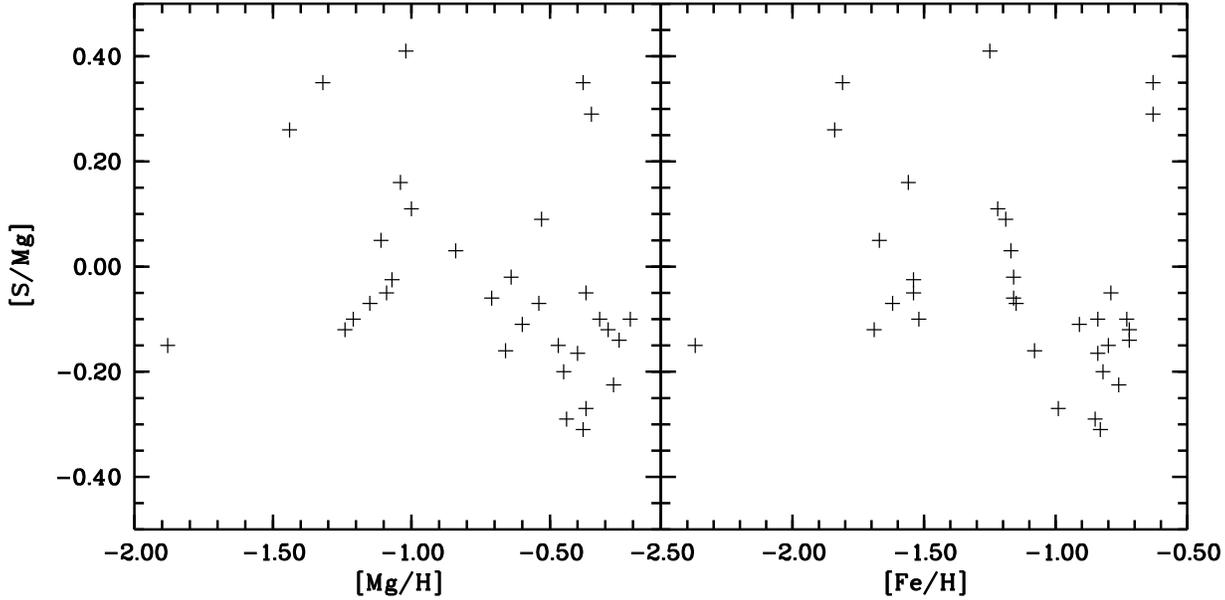,clip=true}
\caption{[S/Mg]  versus [Mg/H] (left panel) and
[Fe/H] (right panel). Mg abundances are from 
\citet{gratton}.}
\label{SMg}
\end{figure*}

\section{Discussion}

\subsection{The behaviour of sulphur versus iron}

In Fig. \ref{sfeall} (bottom panel)
we show the usual [S/Fe] versus [Fe/H]
plot, which displays a very clear rise of [S/Fe]
from solar metallicity up to [Fe/H]$\sim -1$ and then a rather large 
scatter.
It is also interesting to consider 
the plot of [S/H] versus [Fe/H] (Fig. \ref{sfeall}, top panel).
This shows a clear correlation between the
abundances of the two elements. There is, however, a very
clear break in the slope around [Fe/H]$\sim -1$, 
and also in this case the larger scatter at lower
metallicities is obvious.
Another intriguing feature is that,  while at low
metallicities [Mg/Fe] (see Fig. \ref{mgfe_gratton}) 
seems to be essentially flat,
in Fig. \ref{sfeall} for [S/Fe] there is a group
of stars which seems to display
a very clear linear increasing trend, as claimed by
\citet{israelian} and \citet{takeda}, and another group
which seems to display a constant [S/Fe], as found by 
\citet{chen02,chen03,ryde,nissen}.

The increase of [S/Fe] with decreasing metallicity,
already highlighted by previous investigations,
is obvious. However, there are a few features, which
are not obvious when taking each data set separately,
that stand out once all the data are assembled
together as in Fig. \ref{sfeall}:
\begin{enumerate}
\item around [Fe/H]$\sim -1$ there is
a sizeable spread in [Fe/H], clearly larger than at higher
metallicities;
\item for lower metallicities the spread increases
with decreasing [Fe/H]
and there is in fact a hint of bimodality. Some stars have
``high'' [S/Fe], which continues to increase, other stars
have ``low'' [S/Fe] which display a ``plateau''.
\end{enumerate}

\noindent
Before attempting to understand the origin of  these features, it
is necessary to assess if it is
astrophysical or if they are originated by 
observational bias or errors.

For the first  feature  one could presume 
that the metallicity bin around  [Fe/H]$\sim -1$
is the most populated, since stars of this metallicity
are included both in samples consisting mainly 
of metal-poor stars and in samples consisting mainly
of more metal-rich stars. The observational scatter, added
to the different systematics of the different
investigations, may give rise to this excess of scatter.
Although this possibility cannot be ruled out and indeed
most certainly contributes to increase the
existing scatter, we point out
that the feature seems to exist both in our own data
and in the data of \citet{nissen}, albeit in either
sample it is less clear, due to small
number statistics.
We are thus inclined to consider this feature to be of
astrophysical origin.

The second feature is apparent 
in this work for the first time.
It was well known that the works of \citet{israelian}
and \citet{nissen} were in disagreement. The first
displays a steady increase of [S/Fe] with
metallicity and the second a clear ``plateau'',
but researchers in the field were inclined to
believe that either set was plagued by some
undetected systematic effect and thus should be 
discarded.
Very intriguingly, our own results are in good agreement
with the trend of \citet{israelian} and \citet{takeda} for some
stars and with that of \citet{nissen} for some other stars,
giving support to the idea that either at low metallicity
there is a large scatter in S abundances, or there are
two distinct populations, one with 
an  increasing  [S/Fe] with decreasing metallicity
and the other one with a ``plateau''.
A possible cause of concern is that all the data
at low metallicity, where this behaviour becomes apparent,
rely exclusively on the lines of Mult.1 ($\rm 4s-4p ^5S^o-^5 P$)
because the other lines are vanishingly small.
If, for example, these lines were affected by significant NLTE
effects one would expect this to be stronger for
hotter stars, thus causing the observed scatter.
However, this cannot be the case, since in our own sample
the ranges in  temperature and gravity, spanned by the stars
with ``high'' [S/Fe] and ``low'' [S/Fe], are the same.
Again, the temperature range spanned by the stars of 
\citet{israelian} is comparable to that  spanned
by the stars of \citet{nissen}, so stars with
``high'' [S/Fe] do not seem to be systematically
different from stars with ``low'' [S/Fe].  
We checked if there is any dependence of the [S/Fe] ratio
on effective temperature or gravity, in order to 
highlight any possible systematic error, but we 
found none. In Fig. \ref{tvsfe} we show the [S/Fe] ratios
versus \teff for all stars, and there appears to be no trend, 
as could be expected if NLTE effects on the lines
of Mult.1 were important.
In Fig. \ref{badfit}
we show  the fit on the  921.2 nm line of Mult. 1 for 
four of stars from our sample:
all the fits are good and, therefore,
unless we overlooked some systematic effect,
the  deduced abundances should be reliable.

The substantial agreement between our sulphur
abundances and those of \citet{nissen} for the stars
in common militates against the idea of any systematic
difference between the two analysies.

After studying the  works of \citet{nissen}
and \citet{israelian} carefully, we conclude that
the only possible systematic difference
is that \citet{nissen} 
subtracted the telluric lines, while \citet{israelian}
included only stars for which at least one line was measurable.

\begin{figure}
\psfig{figure=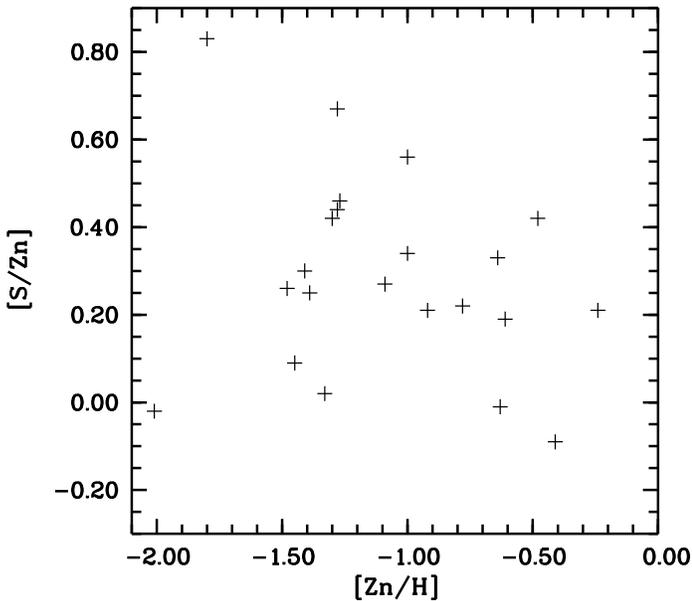,clip=true}
\caption{[S/Zn]  versus [Zn/H]. Zn abundances are from 
\citet{gratton}.}
\label{SZn}
\end{figure}

\begin{figure}
\psfig{figure=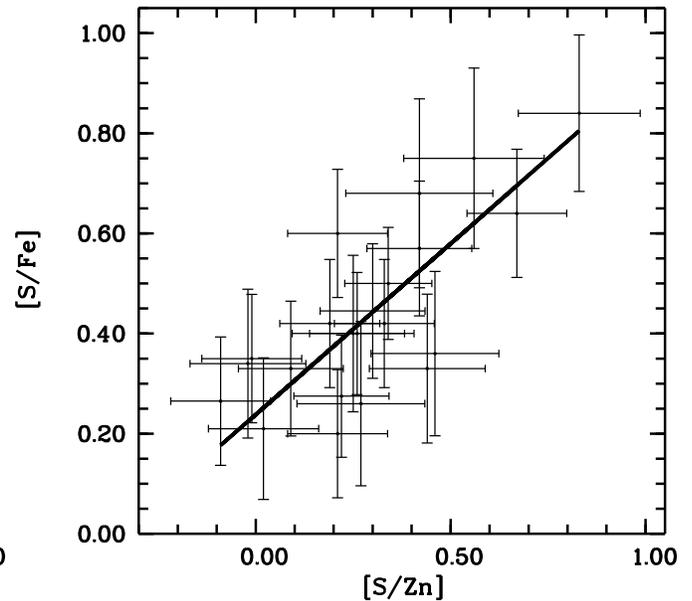,clip=true}
\caption{[S/Fe]  versus [S/Zn]. Zn abundances are from 
\citet{gratton}. The solid line is a least squares
fit to the data:  [S/Fe]=0.68 [S/Zn]+0.24.}
\label{szn_sfe}
\end{figure}

\subsection{Kinematic properties}

In order to classify the stars on the basis of
their kinematics, we computed space velocities 
for all the stars for which parallaxes and proper motions
are available from the Hipparcos or Tycho
catalogues \citep{hipparcos}.
When available radial velocities were taken from our own measurements,
or from
\citet{nord},
or, in sequence,
\citet{latham} or  \citet{Beers}.
When no other data were availabe the radial velocity
in  Simbad was used.
To transform radial velocities and proper motions into
space velocities, we used the transformation matrices
of \citet{johnson_soderblom}, except that we adopted a
left-handed coordinate system with $U$ directed towards 
the Galactic anticentre, $V$ in the direction of Galactic
rotation and $W$ towards the North Galactic Pole.
The kinematic data for the stars are given in Table
\ref{kinetab}.
In Fig. \ref{sfeall} there is a group of 15 stars
with [S/Fe]$>0.48$,
which seem to follow a different trend from the other
stars. We inspected various kinematical properties:
the Toomre velocity ($T=\sqrt{U^2+W^2}$),
the speed ($S=\sqrt{U^2+V^2+W^2}$),
the rotational velocity. However, this group
of stars does not appear to have any
kinematical property distinct from the other ones.

With the velocity and position data,
we integrated
the orbits in the same way as done in \citet{gratton},
and the resulting orbital parameters are reported in 
Table \ref{kinetab}. We used these data to classify
the stars into {\em thin disc}, {\em dissipative
component}, and {\em accretion component} using the
same criteria as \citet{gratton}.
We note four  stars which have apogalactic distances
above 100 kpc and extremely high eccentricities, 
which could be in fact runaway stars
on parabolic orbits, and all four are
classified as belonging to the accretion
component. Not surprisingly, two
of them are the well known extremely metal-poor 
dwarfs \object{G64-12} and \object{G64-37};
the other two are intermediate metallicity
stars \object{HD 105004} and \object{G53-41}.

From Fig. \ref{sfe_pops} we 
see that we find both ``high'' [S/Fe] stars
and ``low'' [S/Fe] stars among the dissipative component,
as well as among the accretion component.
On average the thin disc stars have lower [S/Fe] ratios
and higher metallicities.
The dissipative component also seems to show a larger scatter in 
[S/Fe] ratios than either of the other populations. 
In Fig. \ref{ra_sfe}  [S/Fe] is shown as a function of 
apogalactic distance; and although on average
stars with Rmax larger than 15 kpc appear to 
have higher [S/Fe] ratios, there are some exceptions.
Moreover, there does not seem to be any preferred range
of Rmax for any given range in [S/Fe].
From Fig. \ref{rp_sfe} one may note a very clear trend of increasing
[S/Fe] with decreasing perigalactic distance; on average, the stars which
penetrate closer to the Galactic centre display higher [S/Fe] ratios.
Kendall's $\tau$ test confirms the reality of this 
correlation with a probability close to 1 ($1-prob\sim 4\times 10^{-19}$).
However, if we consider the different populations separately,
we conclude that neither the accretion component nor
the thin disc displays any trend of [S/Fe] versus Rmax,
and it is only the dissipative component which displays
this trend.

Thus the dissipative 
component is in qualitative agreement with the
theoretical results of \citet{fenner}, who predict
that at any metallicity the [S/Fe] ratio should be higher for stars
in the inner disc 
(discussion in Section 9.3). The thin disc, instead, is not
in agreement with this prediction.

\subsection{The behaviour of sulphur with  magnesium and zinc}
 
It is instructive to compare our data for sulphur with
the data for another $\alpha$ element, so we took
magnesium from \citet{gratton}. The situation for silicon
is essentially the same.
The data are displayed in Fig. \ref{mgfe_gratton}.
At first sight the plot seems very similar to the lower
panel of Fig. \ref{sfeall}; however, a closer
inspection reveals that while for Mg for any given
[Fe/H], the [Mg/Fe] value spans a range of 0.3--0.4 dex,
the range can be as large as 0.7 dex for [S/Fe].
In Fig. \ref{mgvsfe} we show [Mg/H] versus [Fe/H]
for 725 Galactic stars from the compilation of \citet{venn04}
as crosses overlaid on [S/H] from the data in Table \ref{kinetab}
(open circles). 
This comparison highlights two differences in the behaviour of the two
elements: 1) the break of slope at [Fe/H]$\sim -1$ is more
pronounced for sulphur than magnesium, the latter displaying a
steeper slope in the high metallicity range ; 2) the width of the
strip in sulphur is wider than in magnesium, in particular 
the eye easily detects a locus of stars with higher
[S/H] for a given [Fe/H], whereas such a locus is not present
for [Mg/H] data.  

For the the sub-sample of our programme stars shared
with \citet{gratton} we may directly form ratios of
sulphur with other elemental abundances, 
since the atmospheric parameters
are the same. In particular,
in Fig. \ref{SMg} we show the ratios [S/Mg] both as 
a function of [Mg/H] (left panel) and [Fe/H] (right panel).
From this plot a large scatter in the S/Mg ratios is apparent,
with a hint of a trend of ratios increasing
with decreasing metallicities. This suggests
that sulphur and magnesium do not vary in lockstep
in the Galactic evolution, although the sample
is too small and the errors too large to reach a
definitive conclusion.

In Fig. \ref{SZn}, instead, we show the [S/Zn] ratios
as a function of [Zn/H].
Morphologically this plot is similar to the one
in Fig. \ref{SMg}, although the metallicity range 
appears more compressed.
This figure  is interesting for many reasons. Zinc 
has been the object of several studies
in Damped Lyman $\alpha$ (DLA) galaxies \citep{pettini94, pettini97,
centurion,nissen}. With respect to other pairs of 
elements, zinc and sulphur have
the advantage that
both are volatile, i.e. form no dust in the warm interstellar medium
\citep{savage_sembach}, so that the nucleosynthetic implications
of the observed ratios may be investigated in DLAs without the need to use
uncertain dust corrections.
There is an ongoing debate on whether the S/Zn may actually
be used as a proxy for $\alpha$/Fe
\citep{centurion} or not \citep{prochaska,fenner}.

From a theoretical point of view, Fe and Zn should behave
differently because Fe is abundantly produced by Type Ia SNe,
while Zn should not \citep{iwamoto}. 
However, \citet{matteucci} have invoked
a Zn production by Type Ia SNe in order
to explain the flatness of the Zn/Fe ratio
in Galactic stars (see \citealt{gratton}).
Without invoking this {\em ad hoc}
production, \citet{fenner} have shown that for a model
of a Milky Way-like galaxy the evolution of the
[S/Fe] ratios is different at different galactocentric radii,
due to the different star formation rates.
However, [S/Zn] should behave in a very similar manner
at all galactocentric radii (see Fig. 7 of \citealt{fenner}).
It would be tempting to interpret the large
scatter in S/Fe ratios 
in our data as the result
of sampling stars which have evolved at different
Galactocentric radii, at least for  the
dissipative component, which shows a clear trend 
of  [S/Fe] with Rmin (see 
Fig. \ref{rp_sfe}).
However, it should be noted that the 
orbits of the dissipative component
display a wide range of eccentricities
and the difference between Rmax and Rmin has
a mean value of $\sim 5$ Kpc. 
Therefore these stars span a significant
range of Galactocentric radii, and it may
not be appropriate to think they represent
the chemical evolution at a ``typical
Galactocentric radius''.
 Note also
that, theoretically, the difference in [S/Fe]
between inner and outer disc ought to be on the 
order of 0.2 dex, while the spread in our data 
is as large as 0.6 dex. 
In Fig. \ref{szn_sfe} we show [S/Fe] versus [S/Zn] for the 22 stars for which
we have measurements of both sulphur and zinc.
There is a clear correlation
between the two, and a linear least squares fit, 
taking errors in both variables
into account, is shown as a
solid line: 
 [S/Fe]=0.68[S/Zn]+0.24.
These data therefore suggest that
[S/Zn] may be used as a proxy of [S/Fe],  
contrary to the
predictions  of the model of \citet{fenner}.
One should, however, keep in mind that this
result rests on sulphur and zinc abundances for only 22 stars.

\section{Conclusions}

In the light of
the current observations, we conclude that 
both stars with  [S/Fe]$\sim 0.4$
and with higher [S/Fe] ratios exist 
at the metal-poor end of the metallicity
distribution of Galactic stars.
We have not been able to ascertain whether
this reflects a larger scatter in sulphur abundances at
low metallicities or  the existence
of two distinct populations.

\balance

\begin{acknowledgements}
EC and PB  
are grateful to Miguel Chavez for useful discussions 
on the topic of stellar abundances  and
for his hospitality at the INAOE, where 
a considerable part of this work was carried out.
We wish to thank Harri Lindgren for providing
unpublished periods for \object{HD 83220} and \object{HD 106516}.
Finally special thanks are due to Gabriella Schiulaz
for helping us to refurbish our language.
\end{acknowledgements}

\bibliographystyle{aa}

\begin{longtable}{lrrrrrrrrrrr}
\caption{Radial velocity, atmospheric parameters and sulphur abundance}\\
\hline\hline
\label{sdat}
 Name    & RV   & Teff & log g& [Fe/H] & Ref &[S/Fe]$_{670}$& [S/Fe]$_{870}$ & [S/Fe]$_{920}$ & [S/Fe] & [S/H] & $\sigma$ \\
         & \kms &  K   & cgs\\
\hline\hline
\endfirsthead
\caption{continued.}\\
\hline\hline
Name    & RV   & Teff & log g& [Fe/H] & Ref &[S/Fe]$_{670}$& [S/Fe]$_{870}$ & [S/Fe]$_{920}$ & [S/Fe] & [S/H] & $\sigma$ \\
        & \kms &  K   & cgs\\
\hline\hline
\endhead
\hline\hline
\endfoot

--09 122 & --47 & 6087 & 4.16 & --1.22 & G03 & $<0.43$ & $  0.17$ &  $ 0.43$ &$ 0.35 $ & $ -0.84$ & 0.11\\
--35 360 &   45 & 5048 & 4.53 & --1.15 & G03 & $<0.59$ & $ <0.59$ &  $ 0.54$ &$ 0.54 $ & $ -0.61$ & 0.15\\
--61 282 &  221 & 5831 & 4.53 & --1.25 & G03 & $<0.69$ & $  0.63$ &  $ 0.66$ &$ 0.64 $ & $ -0.61$ & 0.08\\
--68 74  &  --5 & 5757 & 4.01 & --0.99 & G03 & $ 0.47$ & $  0.36$ &  $ 0.30$ &$ 0.35 $ & $ -0.64$ & 0.08\\
 +02 263 &  --8 & 5754 & 4.87 & --2.17 & Apc & $<1.08$ & $ <1.00$ &  $ 0.91$ &$ 0.91 $ & $ -1.26$ & 0.14\\
--13 482 &   24 & 6194 & 4.34 & --1.61 & G96 & $ - - $ & $  0.33$ &  $ 0.32$ &$ 0.32 $ & $ -1.29$ & 0.08\\
--69 109 &   61 & 5486 & 2.63 & --0.95 & G00 & $ 0.16$ & $  0.15$ &  $ 0.51$ &$ 0.34 $ & $ -0.61$ & 0.06\\
 +09 352 & --64 & 6020 & 4.20 & --2.09 & Apc & $ - - $ & $  - - $ &  $ 0.34$ &$ 0.34 $ & $ -1.75$ & 0.10\\
--60 545 &   11 & 5744 & 4.35 & --0.82 & G03 & $ 0.17$ & $  - - $ &  $ - - $ &$ 0.17 $ & $ -0.65$ & 0.12\\
+10 380  &    6 & 5739 & 4.12 & --0.72 & G03 & $ 0.31$ & $  - - $ &  $ - - $ &$ 0.31 $ & $ -0.41$ & 0.15\\
--09 6150& --33 & 5337 & 4.55 & --0.63 & G03 & $ 0.55$ & $  0.56$ &  $ 0.62$ &$ 0.60 $ & $ -0.03$ & 0.08\\
+20 571  &--117 & 5863 & 4.24 & --0.83 & N97 & $ 0.09$ & $  - - $ &  $ - - $ &$ 0.09 $ & $ -0.74$ & 0.14\\
--47 1087&   11 & 5625 & 4.82 & --0.79 & G03 & $ 0.37$ & $  - - $ &  $ - - $ &$ 0.37 $ & $ -0.42$ & 0.13\\
--03 592 &  120 & 5827 & 4.33 & --0.83 & G03 & $ 0.14$ & $  - - $ &  $ - - $ &$ 0.14 $ & $ -0.69$ & 0.11\\
--26 1453&   90 & 5900 & 4.37 & --0.63 & N97 & $ 0.21$ & $  - - $ &  $ - - $ &$ 0.21 $ & $ -0.42$ & 0.15\\
--57 806 &   55 & 5792 & 4.20 & --0.91 & G03 & $ 0.33$ & $  0.14$ &  $ - - $ &$ 0.20 $ & $ -0.71$ & 0.08\\
--65 253 &   81 & 5351 & 4.57 & --1.52 & G03 & $ - - $ & $ <0.84$ &  $ 0.21$ &$ 0.21 $ & $ -1.31$ & 0.10\\
--27 666 &  111 & 5970 & 4.45 & --1.54 & G03 & $<0.48$ & $ <0.48$ &  $ 0.40$ &$ 0.40 $ & $ -1.14$ & 0.12\\
+05 824  & --15 & 5897 & 4.33 & --1.08 & G03 & $ 0.26$ & $  - - $ &  $ - - $ &$ 0.26 $ & $ -0.82$ & 0.13\\
-59 1024 &  237 & 5894 & 4.49 & --1.69 & G03 & $<0.63$ & $  - - $ &  $ 0.33$ &$ 0.33 $ & $ -1.36$ & 0.09\\
+12 853  &   28 & 5388 & 4.62 & --1.17 & N97 & $<0.26$ & $  - - $ &  $ - - $ &$<0.26 $ & $<-0.91$ & 0.14\\
--33 3337&   71 & 6079 & 4.03 & --1.28 & G03 & $<0.62$ & $  - - $ &  $ - - $ &$<0.62 $ & $<-0.62$ & 0.12\\
--57 1633&  260 & 6013 & 4.34 & --0.84 & G03 & $<0.05$ & $  - - $ &  $ - - $ &$<0.05 $ & $< 0.05$ & 0.12\\
--45 3283&  316 & 5692 & 4.82 & --0.85 & G03 & $<0.49$ & $  - - $ &  $ - - $ &$<0.49 $ & $<-0.36$ & 0.14\\
 G 88--40& --35 & 5967 & 4.26 & --0.80 & N97 & $ 0.18$ & $  - - $ &  $ - - $ &$ 0.18 $ & $ -0.62$ & 0.16\\
--15 2656&  117 & 5923 & 4.14 & --0.85 & G03 & $ 0.12$ & $  - - $ &  $ - - $ &$ 0.12 $ & $ -0.73$ & 0.10\\
 G 46--31&  218 & 6021 & 4.44 & --0.75 & N97 & $<0.21$ & $  - - $ &  $ - - $ &$<0.21 $ & $<-0.54$ & 0.18\\
--20 3540&  167 & 6029 & 4.32 & --0.79 & N97 & $ 0.05$ & $  - - $ &  $ - - $ &$ 0.05 $ & $ -0.74$ & 0.12\\
--25 9024&  120 & 5831 & 4.36 & --0.80 & N97 & $ 0.19$ & $  - - $ &  $ - - $ &$ 0.19 $ & $ -0.61$ & 0.13\\
 G 12--21&  100 & 6013 & 4.44 & --1.27 & G03 & $<0.68$ & $  - - $ &  $ - - $ &$<0.68 $ & $<-1.00$ & 0.13\\
--09 3468&  --5 & 6232 & 4.29 & --0.72 & G03 & $ 0.33$ & $  - - $ &  $ - - $ &$ 0.33 $ & $ -0.39$ & 0.12\\
+02 2538 &  155 & 6133 & 4.41 & --1.69 & G03 & $<0.56$ & $ <0.18$ &  $ - - $ &$<0.18 $ & $<-1.55$ & 0.09\\
--37 8363&  226 & 5543 & 3.88 & --0.70 & G03 & $<0.34$ & $  - - $ &  $ - - $ &$<0.34 $ & $<-0.36$ & 0.28\\
--38 8457&  145 & 5964 & 4.32 & --1.84 & G03 & $ - - $ & $  0.66$ &  $ - - $ &$ 0.66 $ & $ -1.18$ & 0.13\\
--56 5169&   13 & 5383 & 4.57 & --0.94 & G03 & $<0.58$ & $ <0.36$ &  $ - - $ &$<0.36 $ & $<-0.58$ & 0.09\\
--45 8786&  245 & 5686 & 4.40 & --0.76 & N97 & $ 0.29$ & $  0.24$ &  $ - - $ &$ 0.26 $ & $ -0.50$ & 0.08\\
--17 4092& --46 & 5574 & 4.55 & --1.14 & G03 & $<0.41$ & $ <0.45$ &  $ - - $ &$<0.41 $ & $<-0.73$ & 0.08\\
--21 4009&  176 & 5541 & 3.79 & --1.67 & G03 & $ 0.63$ & $  0.59$ &  $ - - $ &$ 0.61 $ & $ -1.06$ & 0.07\\
+04 2969 &   18 & 5850 & 3.95 & --0.84 & G03 & $ 0.25$ & $ 0.30 $ &  $ - - $ &$ 0.28 $ & $ -0.56$ & 0.07\\
--15 4042&  310 & 4996 & 4.65 & --1.38 & G03 & $<0.39$ & $ <0.72$ &  $ - - $ &$<0.39 $ & $<-0.99$ & 0.11\\
--57 6303&    9 & 4869 & 4.62 & --1.39 & G03 & $<0.83$ & $  - - $ &  $ - - $ &$<0.83 $ & $<-0.56$ & 0.17\\
+06 3455 &--136 & 5713 & 4.35 & --0.84 & G03 & $ 0.34$ & $  0.33$ &  $ 0.47$ &$ 0.42 $ & $ -0.42$ & 0.08\\
 +02 3375&--380 & 6018 & 4.20 & --2.37 & G96 & $ - - $ & $  - - $ &  $ 0.34$ &$ 0.34 $ & $ -2.03$ & 0.11\\
 +05 3640& --1  & 5023 & 4.61 & --1.19 & G03 & $<0.93$ & $ <0.82$ &  $ 0.75$ &$ 0.75 $ & $ -0.44$ & 0.15\\
--59 6824& --47 & 6070 & 4.17 & --1.54 & G03 & $<0.63$ & $  0.49$ &  $ 0.40$ &$ 0.44 $ & $ -1.11$ & 0.09\\
+13 3683 &   86 & 5726 & 3.78 & --2.43 & Apc & $ - - $ & $  - - $ &  $ 0.47$ &$ 0.47 $ & $ -1.96$ & 0.10\\
G 21--22 &   60 & 6123 & 4.28 & --0.88 & Apc & $<-0.01$& $<-0.51$ &  $-0.14$ &$-0.14 $ & $ -1.02$ & 0.11\\
--45 13178&  30 & 5968 & 4.40 & --1.81 & G03 & $- -  $ & $  0.84$ &  $ - - $ &$ 0.84 $ & $ -0.97$ & 0.12\\
 +10 4091&--193 & 5503 & 4.55 & --1.45 & G03 & $<0.55$ & $ <0.15$ &  $ - - $ &$<0.15 $ & $<-1.30$ & 0.09\\
--12 5613& --14 & 5668 & 3.79 & --1.18 & G03 & $<0.42$ & $ <0.40$ &  $ - - $ &$<0.40 $ & $<-0.78$ & 0.10\\
--21 5703&--173 & 5779 & 4.54 & --1.09 & G03 & $ 0.33$ & $  0.37$ &  $ - - $ &$ 0.35 $ & $ -0.74$ & 0.09\\
+09 4529 &--248 & 6023 & 4.31 & --1.17 & G03 & $<0.23$ & $  0.36$ &  $ - - $ &$ 0.36 $ & $ -0.81$ & 0.13\\
--19 5889& --34 & 5893 & 4.12 & --1.16 & G03 & $ 0.36$ & $  0.44$ &  $ 0.56$ &$ 0.50 $ & $ -0.66$ & 0.05\\
+04 4551 &--117 & 5892 & 4.14 & --1.40 & G96 & $<0.54$ & $ <0.54$ &  $ 0.46$ &$ 0.46 $ & $ -0.94$ & 0.14\\
 +04 4674& --84 & 5772 & 4.03 & --0.73 & G03 & $ 0.24$ & $ 0.26 $ &  $ 0.56$ &$ 0.42 $ & $ -0.31$ & 0.08\\
--28 17381&--105& 5810 & 4.50 & --1.16 & G03 & $<0.50$ & $ <0.20$ &  $ 0.39$ &$ 0.39 $ & $ -0.77$ & 0.13\\
 +17 4708&--295 & 6016 & 4.04 & --1.62 & Apc & $ 0.42$ & $  0.38$ &  $ 0.40$ &$ 0.40 $ & $ -1.22$ & 0.07\\
+07 4841 &--232 & 5980 & 4.00 & --1.59 & R88-P93 & $ 0.51 $&$  0.53$ &$ 0.56$ &$ 0.54 $& $ -1.05$ & 0.08\\
HR 8515  &   31 & 5211 & 3.36 & --1.56 & G03 & $<0.75$ & $ <0.70$ &  $ 0.68$ &$ 0.68 $ & $ -0.88$ & 0.16* \\
 G 18--54&--217 & 5878 & 3.93 & --1.33 & Apc-P93 & $ <0.07 $&$<-0.01$ &  $-0.03$ &$-0.03 $ & $ -1.36$ & 0.14*  \\
--09 6149& --30 & 5756 & 4.26 & --0.63 & G03 & $0.45 $ & $  0.44$ &  $ 0.69$ &$ 0.57 $ & $ -0.06$ & 0.09\\
 GD 660  & --69 & 5712 & 4.50 & --1.64 & Apc & $<1.09$ & $ <1.16$ &  $ 0.40$ &$ 0.40 $ & $ -1.24$ & 0.17\\
GCRV 7547& --39 & 6272 & 4.03 & --0.42 & N97 & $ 0.04$ & $  - - $ &  $ - - $ &$ 0.04 $ & $ -0.38$ & 0.14\\
G 75 031 &   57 & 5884 & 4.24 & --1.25 & Apc & $ 0.52$ & $  - - $ &  $ - - $ &$ 0.52 $ & $ -0.73$ & 0.14\\
--48 4818& --15 & 6503 & 4.11 & --0.43 & G03 & $>-0.18$& $  - - $ &  $ - - $ &$>-0.18$ & $>-0.61$ & 0.18\\
--21 3420&    6 & 5946 & 4.41 & --1.04 & N97 & $<0.45 $& $  - - $ &  $ - - $ &$ <0.45$ & $<-0.59$ & 0.14\\
 \\                                                                      
\hline\hline
\\
\multispan{11}{The column (``Ref'') indicates the source of atmospheric parameter:\hfill}\\
\multispan{11}{G03: \citet{gratton}\hfill}\\
\multispan{11}{N97: \citet{NS}\hfill}\\
\multispan{11}{P93: \citet{pila}\hfill}\\
\multispan{11}{Apc: Alonso, private comunication\hfill}\\
\multispan{11}{* are SB2 stars. see Appendix A\hfill}\\

\end{longtable}

\Online

\appendix

\section{Remarks on individual stars}

\begin{enumerate}

\item
\object{HD 3567}:
The difference between  the determination of sulphur abundance 
from the lines of Mult. 6 and that from those of Mult. 1 
is not explained; it may be due to weakness
of the Mult. 6 lines.

\item
\object{HD 17072}:
The star is an RHB according to \citet{carney}.  \citet{gratton98}
notes that it may be on the first ascent of the giant branch rather than
on the horizontal branch.
The incoherent sulphur abundance in the three regions remains unexplained.
For Mult. 1 the good fit is based on two lines, while the third one was
discarded for its contamination by a telluric line; the slight contamination
by telluric of the two considered lines was easily removed.
Sulphur abundance for 
Mult. 8 and 6 is based on one sulphur line for each multiplet.
We note that a broadening higher than the instrumental resolution is required;
the microturbulence of 2.1 \kms from \citet{carney} was adopted.

\item
\object{G  76 --21}:
It has peculiar line profiles. 
The lines appear broad (higher than instrumental resolution)
with a flat and double core (see Fig. \ref{doppiag76}).
\citet{carney} suspected it of being a double--lined system; the duplicity is 
not confirmed by \citet{latham}, who measured 
only a small amplitude variation
for radial velocity.

\begin{figure*}
\psfig{figure=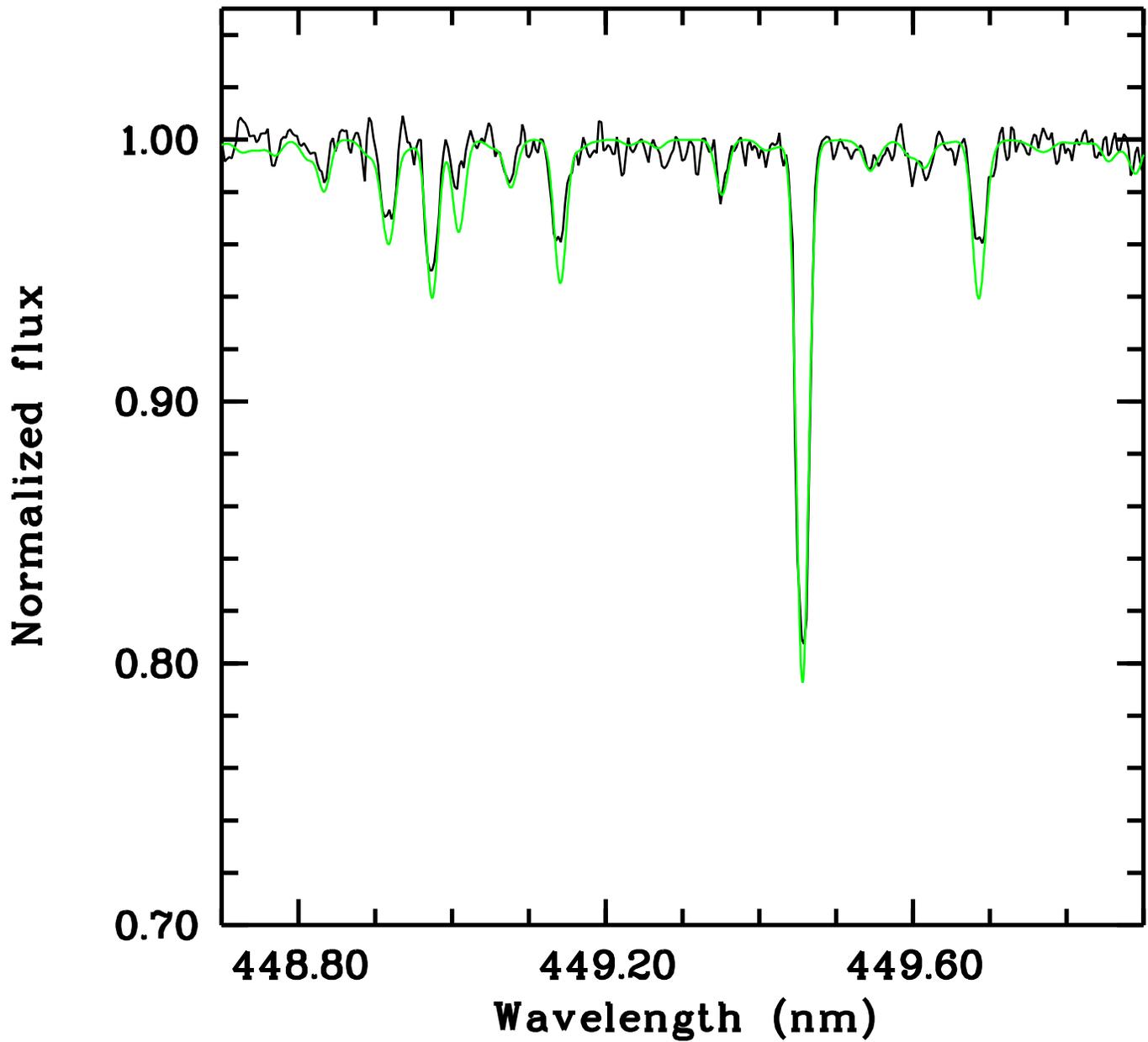,width=\hsize,clip=true}
\caption{\object{G 76 --21} is suspected of being a double-lined system.
In fact, lines:
\ion{Fe}{ii} 448.9183nm, \ion{Fe}{ii} 449.1405 nm, and 
\ion{Cr}{i} 449.6852 nm
show a flat and double core.
}
\label{doppiag76}
\end{figure*}

\item
\object{HD 83220}:
This star presents broad lines. The rotational velocity (9 \kms) of 
\cite{nord} is the highest in the sample of stars in common with ours.
This star is a spetroscopic binary
for which Lindgren (private communication) determined a
preliminary orbit with a period of 765.8 days.

\item
\object{HD 103723} and \object{HD 105004}:
The low $\alpha$--enhancement found by 
\cite{nissen}
is not evident from our spectra.
Both stars are suspected binares from Hipparcos data:
\object{HD 103723} is D (duplicity induced variability);
\object{HD 105004} is S (suspected not single) and X (probably an
astrometric binary with short period).

\item
\object{HD 106038}:
We suspect the star is double. We recall the peculiar abundances
found by \cite{NS}, but not by
\cite{chen01}.

\item
\object{HD 106516}:
A spectroscopic binary. \citet{latham} give a period of
853.2 d, and 
Lindgren (private comunication) a period of 841 d.
It is one of the few stars presenting 
lines broader than instrumental resolution
in   agreement with \cite{nord} who give $v\sin i =8$ \kms.
\cite{peterson} establish it as a halo blue straggler, which
had already been suggested by 
\cite{fuhr}  based on 
the unexpectedely high projected rotational velocity.
Its blue straggler nature may in fact explain the
absence of detectable Be in its atmosphere \citep{molaro}.

\item
\object{HD 113083}:
We observe the duplicity discovered by
\citet{lind} and confirmed by
\citet{NS} and by \citet{nissen00}.
\object{HD 113083} is an SB2 with nearly identical sets of lines;
see \citet{NS} for the resolved system parameters and abundances.

\item
\object{HD 132475}:
We note the disagreement between the temperature adopted by us (5541 K)
and the one adopted by \citet{ryde} (5810 K).
The synthetic spectrum computed with the temperature adopted by us fits
the observed H$_\alpha$ better than  the one  computed with the \teff 
used  by \citet{ryde}.

\item
\object{HD 204155}:
We note a large difference of the value of the sulphur abundance
from Mult. 8 and 6 with
respect to  that derived from Mult. 1.
No signs of duplicity can be deduced from the 
radial velocity data of \citet{latham}
covering more than 3000 days.

\item
\object{HD 211998}:
Only the 921.28 nm line of Mult. 1 was
detected.
The H$_{\alpha}$ profile is not in good agreement with the
synthetic spectrum.
In spite of being a much studied bright star (V=5.29),
its duplicity is questionable.
\citet{malaroda73} and
\citet{malaroda75}
classified it as a spectroscopic binary composed of an A and an F star.
\citet{gray} suspected it of having a composite spectrum.
However, \citet{lammc}
discarded the duplicity hypothesis.

The presence of Li \citep{maurice} but not Be
\citep{molaro},
is at odds with predictions of standard stellar evolution theory.

Hipparcos \citep{hipparcos} did not detect any companion and the 
speckle  measurements \citep{fourth} 
are uncertain.

\item
\object{G 18--54}: 
Sulphur was  detected only for Mult. 1.
We neglected the contribution by the fainter companion.
Weaker red--shifted lines
from a fainter companion are present in our spectra.
\cite{carney} classified this star as SB2, and
\cite{nissen} confirmed it.
\cite{latham} computed the period of the orbit
(P=493.00 d).

\item
\object{HD 219175}:
For this star we find different temperature determinations:
T$_{eff}$ (adopted) = 5756 K\citep{gratton}; T$_{eff}$(B--V) = 5844 K; 
T$_{eff}$(H$_\alpha$) = 5856 K \citep{gratton}.
From the fit of H$_\alpha$ wings, we obtained T$_{eff}$(H$_\alpha$) = 6050 K.
Moreover, many metal line cores are flat and their profiles are
slightly asymmetric. This fact is evident in sulphur lines (see for
example Fig. \ref{s_hd219175}).
From these facts we suspect this star of being double.

\begin{figure*}
\psfig{figure=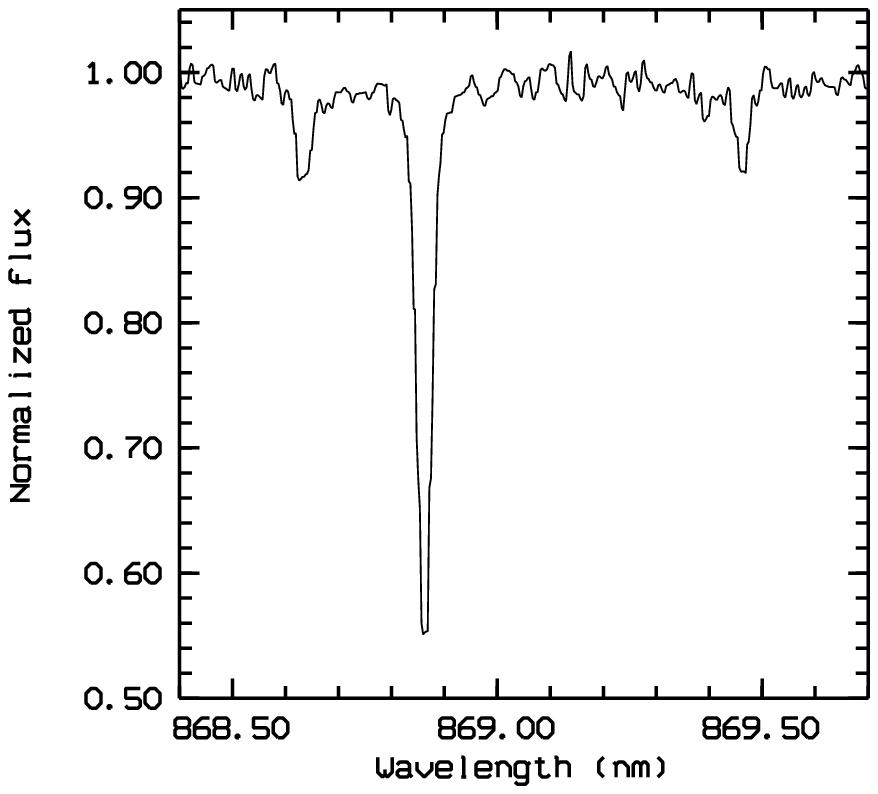,width=\hsize,clip=true}
\caption{\object{HD 219175} shows a flat core of 
\ion{Si}{i} 868.6352 nm, \ion{Fe}{i} 868.8624 nm and 
\ion{S}{i} 869.3931 nm and 869.4626 nm.
}
\label{s_hd219175}
\end{figure*}

\end{enumerate}

{\scriptsize
\begin{longtable}{rrrrrrrrrl}
\caption{Observational data}\\
\hline\hline
\label{data}
   HD    &Rem &     BD/CD&  G      & HIP   &$\alpha$(2000)& $\delta$(2000)& JD--2400000.5   & S/N & Telescope\\
\hline\hline
\endfirsthead
\caption{continued.}\\
\hline\hline
   HD    &Rem &     BD/CD&  G      & HIP   &$\alpha$(2000)& $\delta$(2000)& JD--2400000.5   & S/N & Telescope\\
\hline\hline
\endhead
\hline\hline
\endfoot
\object{  3567  } &    &\object{ --09 122} &\object{ 270--2}3 &  3026 & 00 38 32     & --08 18 33   &  50085.050970000 & 135 & ESONTTB\\
                  &    &                   &                  &       &              &              &  52152.248250600 & 240 & ESO-VLT-U2\\
                  &    &\object{ --35 0360}&\object{ 269--87} &  5004 & 01 04 06     & --34 40 29   &  51794.410460600 & 160 & ESO-VLT-U2\\
                  &    &                   &\object{ 268--32} &  3446 & 00 44 04     & --13 55 26   &  52151.407573600 & 130 & ESO-VLT-U2\\
                  &    &\object{--61 282}  &                  &  7459 & 01 36 06     & --61 05 03   &  50085.077490000 & 130 & ESONTTB\\
                  &    &                   &                  &       &              &              &  52149.299538900 & 250 & ESO-VLT-U2\\
\object{  10607 } &    &\object{--68 74 }  &                  & 7869  & 01 41 15     & --67 40 37   &  51795.410954400 & 190 & ESO-VLT-U2\\
                  &    &\object{ +02 263 } &\object{ 71--33 } &       & 01 45 14     &  +03 30.8    &  52150.425270600 & 210 & ESO-VLT-U2\\
\hline\hline
\end{longtable}
}

{\scriptsize
\tabskip=0pt
\begin{longtable}{rrrrrrrrrrrrrrrr}
\caption{Observational data}\\
\hline\hline
\label{kinetab}
name         &  $\pi$ & RV     &    U    &  V       & W       & \teff & log g  & [Fe/H]  &[S/Fe]& Ref$^a$
 &  Rmin    &  Rmax     &  Zmax   & e    & pop \\        
     &                mas  &  \multispan4{\hfill \kms    \phantom{mm}  }   &  K    & cgs    & dex     & dex
 & &  kpc     & kpc       & kpc    \\
\\
\hline\hline
\endfirsthead
\caption{continued.}\\
\hline\hline
 name         &  $\pi$ & RV     &    U    &  V       & W       & \teff & log g  & [Fe/H]  &[S/Fe]& Ref$^a$        
 &  Rmin    &  Rmax     &  Zmax   & e    & pop \\        
     &                mas   &  \multispan4{\hfill \kms  \phantom{mm}  }      &  K    & cgs    & dex     & dex
 &  &  kpc     & kpc       & kpc    \\
\hline\hline
\endhead
\hline\hline
\endfoot
$\object{ -09\ 122  } $&  9.57  &$ -47  $&$ -136  $&$  -235 $ &$  -44$ &$ 6087$& $ 4.16$&$ -1.22$& $ +0.33$ & 1       &   0.141  &  10.612   &  7.615 & 0.974 & 3\\
$\object{ -35~0360  } $& 16.28  &$ +45  $&$ +110  $&$  -176 $ &$  -26$ &$ 5048$& $ 4.53$&$ -1.15$& $ +0.54$ & 1       &   0.926  &   9.488   &  0.379 & 0.822 & 2\\
$\object{  -61~0282 } $& 11.63  &$+221  $&$ -238  $&$  -260 $ &$  -37$ &$ 5831$& $ 4.53$&$ -1.25$& $ +0.64$ & 1       &   0.554  &  16.694   &  7.748 & 0.936 & 3\\
$\object{  -68\ 74  } $& 14.01  &$  -5  $&$   -4  $&$  -143 $ &$ +121$ &$ 5757$& $ 4.01$&$ -0.99$& $ +0.35$ & 1       &   2.430  &   8.511   &  4.838 & 0.556 & 2\\
$\object{  02~263   } $&     *  &$  -8  $&$    *  $&$     * $ &$    *$ &$ 5754$& $ 4.87$&$ -2.17$& $ +0.91$ & 1       &       *  &       *   &      * &     * & 3\\
$\object{  -13\ 482 } $&  8.67  &$ +24  $&$  -29  $&$   -99 $ &$  -37$ &$ 6194$& $ 4.34$&$ -1.61$& $ +0.32$ & 1       &   3.447  &   8.761   &  0.446 & 0.435 & 2\\
$\object{ -69\ 109   } $&  7.57  &$ +61  $&$   +5  $&$   -70 $ &$  -22$ &$ 5486$& $ 2.63$&$ -0.95$& $ +0.34$ & 1       &   4.675  &   8.476   &  0.206 & 0.289 & 2\\
$\object{  +09\ 352 } $&  5.22  &$ -64  $&$ +126  $&$  -207 $ &$ +150$ &$ 6020$& $ 4.20$&$ -2.09$& $ +0.34$ & 1       &   0.505  &  10.789   & 10.353 & 0.910 & 3\\
\end{longtable}
}

\end{document}